\documentclass[aps, pra, floatfix]{revtex4}
\usepackage{graphicx}
\usepackage{amsmath}
\usepackage{setspace}

\begin{document}
% \linenumbers
%-----------------------TITLE, AUTHORS & AFFILIATIONS---------------------------
\title{New insights into the semiclassical Wigner \\  treatment of photodissociation dynamics}

\author{$^{1,2}$W. Arbelo-Gonz\'alez, $^{1}$L. Bonnet{\footnote{Corresponding author. Email: l.bonnet@ism.u-bordeaux1.fr}} and 
$^{3}$A. Garc\'ia-Vela}

\affiliation{$^{1}$CNRS, Univ. Bordeaux, ISM, UMR 5255, 33405, Talence, France \\
$^{2}$Departamento de F\'{\i}sica General, Instituto Superior de
Tecnolog\'{\i}as y Ciencias Aplicadas, Habana 6163, Cuba\\
$^{3}$Instituto de F\'{\i}sica Fundamental, C.S.I.C., Serrano 123,
28006 Madrid, Spain}
\date{\today}
%----------------------------------ABSTRACT-------------------------------------
\begin{abstract}
\noindent The \emph{semiclassical Wigner treatment} of Brown and Heller [J. Chem. Phys. 75, 186 (1981)] is applied to triatomic
direct photodissociations with the aim of accurately predicting final state distributions at relatively low computational cost,
and having available a powerful interpretative tool. For the first time, the treatment is full-dimensional. The proposed
formulation closely parallels the quantum description as far as possible. An approximate version is proposed, which is still
accurate while numerically much more efficient. In addition to be weighted by usual vibrational Wigner distributions, final
phase space states appear to be weighted by new \emph{rotational Wigner distributions}. 
%which do not seem to have been used before. 
%which do not seem to have been considered before. 
These densities have remarkable structures clearly showing that classical trajectories most contributing to rotational state
$j$ are those reaching the products with a rotational angular momentum close to $[j(j+1)]^{1/2}$ (in $\hbar$ unit). The
previous methods involve running trajectories from the reagent molecule onto the products. The alternative
\emph{backward approach} [L. Bonnet, J. Chem. Phys. 133, 174108 (2010)], in which trajectories are run in the reverse
direction, is shown to strongly improve the numerical efficiency of the most rigorous method in addition to be
\emph{state-selective}, and thus, ideally suited to the description of state-correlated distributions measured in velocity
imaging experiments. The results obtained by means of the previous methods are compared with rigorous quantum results in the
case of Guo's 
%reduced dimensionality 
triatomic-like model of methyl iodide photodissociation [J. Chem. Phys. 96, 6629 (1992)] and an astonishing agreement is found. 
In comparison, the standard method of Goursaud \emph{et al.} [J. Chem. Phys. 65, 5453 (1976)] is only semi-quantitative.
%It is found that the vibrational and rotational Wigner densities used to weight final phase space states may play a crucial
%role that standard or Gaussian bins cannot play. \\
%and good agreement is found. The standard and Gaussian binning procedures are also applied, leading to less accurate results. In %particular, they
%do not allow to systematically reproduce inverted vibrational or rotational state populations, contrary to the semiclassical Wigner %treatment. This finding is explained from the structure of reagent and product Wigner densities.
%The contribution of final
%phase space states to final quantum states is also treated by means of the standard and Gaussian binning (GB) procedures.

\end{abstract}
\maketitle
%--------------------------------INTRODUCTION-----------------------------------
\section{Introduction}
\label{sec:introduction}

Modern molecular beam and spectroscopic technics allow the measurement of quantum state distributions of photodissociation 
products with an amazing level of precision \cite{Suits1,Suits2,Chiche,Parker1,Parker2}. Accurate theoretical descriptions of
the mechanics of molecular fragmentation \cite{Schinke} are thus needed to reproduce and rationalize these data, or predict
them when experiments cannot be performed. Beyond their fundamental interest, photodissociation dynamics studies provide very 
useful data to specialists of planetary atmospheres \cite{Planet} or interstellar clouds \cite{Astro}, and they form a branch
of molecular physics which has continuously generated technological progress for more than a century. 

The goal of the present paper is to report some new insights into the \emph{semiclassical Wigner treatment}
\cite{Heller2,Schinke2} of the dynamics of direct triatomic photodissociations (see also refs.~\cite{Heller1,Schinke,Walker,Koy,Goursaud,Truhlar,Part1,Part2,Part3} for related works). In these very common processes,
nascent products strongly repel each other after the optical excitation and are completely free in a few tenths of femtoseconds \cite{Schinke}. The practical interest of the semiclassical Wigner method, proposed about three decades ago by Brown and Heller \cite{Heller2}, is that it was found to be more accurate than the earlier and more classical approach of Goursaud \emph{et al.} \cite{Heller2,Goursaud}, and leads to calculations much easier to perform than exact quantum calculations \cite{Balint1,Balint2,Balint3,Balint4,Octavio1,Octavio2,Momo1,Momo2,Alberto1,Alberto2,Manthe,Wang,Hua1}. 
As far as polyatomic molecules are concerned, the latter are usually prohibitive and the semiclassical Wigner treatment might
thus be an interesting alternative. From the fundamental side, the interpretative power of this approach makes it a powerful
tool for rationalizing the dynamics.
%as will be shown later in this report.

The semiclassical Wigner method is based on the notion of \emph{Wigner distribution} associated with a given quantum state \cite{Wigner,Heller3,LeeScully,Berry,Bogota,Lee,Stanek,Case,Schinke}. For a one-dimensional system of configuration coordinate
$x$ and conjugate momentum $p_x$ in the state $|\Psi\rangle$, the Wigner distribution is defined as 
\\
\begin{equation}
 \rho(x,p_x) = \frac{1}{\pi \hbar}\int\;ds \; 
 e^{2ip_xs/\hbar}\;
 \langle \Psi|x+s \rangle\;
 \langle x-s|\Psi \rangle.
 \label{I1}
\end{equation}
\\
This fascinating density was introduced by Wigner in 1932 \cite{Wigner} as a convenient tool for quantum mechanically
correcting the Gibbs-Boltzmann theory of thermodynamic equilibrium in the limit of small temperatures. Integration of
$\rho(x,p_x)$ over $p_x$ leads to $|\langle x|\Psi \rangle|^2$ while integration over $x$ leads to
$|\langle p_x|\Psi \rangle|^2$. In addition to that, calling respectively $H$ and $\hat{H}$ the classical and quantum
Hamiltonians of the system, integration of $\rho(x,p_x)H$ with respect to $x$ and $p_x$ leads to
$\langle \Psi|\hat{H}|\Psi \rangle$. One might thus be tempted to call $\rho(x,p_x)$ the phase space distribution
corresponding to $|\Psi\rangle$. \emph{Stricto sensu}, however, one should not, for $\rho(x,p_x)$ takes negative values in some
parts of the phase space when $\langle x|\Psi \rangle$ (or $\langle p_x|\Psi \rangle$) involves at least one node, and the
uncertainty principle puts a shadow on the phase space concept. Nevertheless, one can formally use this non conventional
density of probability as an usual one.

The beauty of the semiclassical Wigner treatment is that it mixes the quantum and classical descriptions in a very natural way \cite{Heller2,Schinke2}. The underlying principle of the treatment is as follows. The population of a given product quantum
state is shown to be proportional to the square modulous of the overlap between (i), the product state wave function, and (ii),
the time-evolved wave packet resulting from the propagation of the initial reagent state (multiplied by the transition moment) 
over a period of time large enough for the whole wave packet to be located in the product channel \cite{Schinke,Heller2}. This
expression is then transformed into an overlap between the Wigner distribution corresponding to the product state, and the
Wigner distribution associated with the time-evolved wave packet \cite{Heller2}. Everything is rigorous up to now. What makes
the Wigner treatment semiclassical is that the Wigner distribution of the time-evolved wave packet is obtained by propagating 
in time the Wigner density of the initial reagent state according to the laws of classical mechanics \cite{Heller2}, just as if
this density was a solution of Liouville equation \cite{Gold}. To recap, initial and final dynamical states (phase space
points) are assigned statistical weights according to quantum mechanics through Wigner distributions while nuclear dynamics are
ruled by classical mechanics. The corresponding mathematical developments are given in section~\ref{II.A}. In this approach,
trajectories are run forward in time from the reagent molecule onto the separated products. We shall call it \emph{forward I}.

%In the very few applications of the semiclassical Wigner method published to date \cite{Heller2,Schinke2}, 
In the only two applications of the semiclassical Wigner method that seem to have been published to date
\cite{Heller2,Schinke2}, rotation motions were frozen. In the present work, however, we take into account the
full-dimensionality of the triatomic system, making thereby the method applicable to processes taking place in realistic
conditions.

In addition to the usual vibrational Wigner distributions \cite{Heller2,Heller1,Schinke2,Truhlar,LeeScully,Lee,Stanek,Case}, 
the method involves \emph{rotational Wigner distributions} which, to our knowledge, are introduced for the first time in the
context of reaction dynamics. These quantities are respectively discussed in sections~\ref{II.B} and~\ref{II.C}. Rotational
Wigner distributions have remarkable structures clearly showing that the classical trajectories most contributing to rotational
state $j$ are those reaching the products with a rotational angular momentum close to $[j(j+1)]^{1/2}$ (in $\hbar$ unit).

The study of a Franck-Condon model process in section~\ref{II.D} allows to illustrate how the \emph{degrading effect}, 
discussed by Gray and Truhlar \cite{Truhlar} and Schinke \cite{Schinke2}, alters state-resolved cross sections en route to
products. This suggests a slight formal modification of the forward I method in order to greatly improve its accuracy. 

Nevertheless, this method turns out to have a limited numerical efficiency. An approximation is thus proposed in
section~\ref{II.E} in order to strongly increase it. The resulting method will be called \emph{forward II}. 
%As a by-product of this approach, we examine in its light the theoretical 
%grounds of the formulation of 
%Goursaud \emph{et al.} \cite{Goursaud}, such as implemented by Schinke \cite{Schinke}. 

%However, final dynamical states are still weighted by the product Wigner distribution 
%whereas in the method of Goursaud \emph{et al.}, the standard binning procedure is used \cite{Part3}. 
%The resulting
%approach leads to very satisfying predictions, 
%as shown in section~\ref{IV}, and is numerically 
%much more efficient than the more rigorous forward I method. 

The recent \emph{backward approach} \cite{Part1,Part2,Part3}, in which trajectories are run from the products onto the reagent
molecule, is shown in section~\ref{III} to be strictly equivalent to the forward I method while being numerically much more
efficient. In addition to that, it is \emph{state-selective}, and thus, ideally suited to the description of state-correlated
distributions measured in velocity imaging experiments \cite{Suits1,Suits2,Chiche,Parker1,Parker2}.

Since we shall also apply the standard method of Goursaud \emph{et al.} \cite{Goursaud}, we briefly recall its main lines in section~\ref{IV}. 

The results obtained by means of the two forward methods and the backward one are compared in section~\ref{V} with rigorous
quantum results \cite{Alberto1,Alberto2} in the case of the reduced dimensionality model of methyl iodide photodissociation of
Guo \cite{Guo2}, and very good - often quantitative - agreement is found. In comparison, the method of Goursaud \emph{et al.}
\cite{Goursaud} is only semi-quantitative, at least for one of the two excited electronic states involved in the process,
within which the system has a more quantum-like behavior. In particular, this approach does not allow to systematically
reproduce inverted vibrational or rotational state populations, contrary to the semiclassical Wigner treatment. 
%This finding is explained in terms of structural 
%differences between product Wigner densities and standard bins.
Section~\ref{VI} concludes.

\section{Forward semiclassical Wigner approach}
\label{II}

\subsection{Accurate formulation}
\label{II.A}

\subsubsection{System}
\label{II.A.1}

Let us consider a triatomic molecule ABC optically excited by a photon of energy $h\nu$ from 
its rovibronic ground state at energy $E_0$ up to a given repulsive electronic excited state.
The future products, say AB and C, strongly repel each other immediately after the photon absorption and 
are formed in a few tenths of femtoseconds. The dissociation is thus direct. The total energy of ABC is $E = E_0 + h\nu$.
The optical excitation is supposed not to excite the rotation motion. Within the framework of this reasonable approximation,
the total angular momentum $J$ is kept at 0 throughout the whole fragmentation process (see refs.~\cite{Schinke,Balint1,Octavio1,Octavio2,
Balint3} for rigorous treatments of the possible rotational transitions implied by an optical excitation). 
$\mathbf{R}$ is the vector going from the 
center-of-mass G of AB to C and $\mathbf{r}$ is the one from A to B. $V_g$ and $V_e$ are the potential energies in the 
ground and excited electronic states, respectively. They both depend on the moduli $R$ and $r$ of $\mathbf{R}$ and $\mathbf{r}$ 
and the angle $\theta$ between them. Far in the asymptotic channel, $V_e$ reduces to the potential energy of the free AB diatom,
denoted $v_e(r)$. The zero of energy is defined as the minimum of $V_e$, or $v_e(r)$, in the separated products. 
$m$ and $\mu$ are, respectively, the reduced masses of AB, and C with respect to AB.
$\bf{d}\equiv\bf{d}(\mathbf{R},\mathbf{r})$ is the transition dipole vector
responsible for the electronic transition \cite{Schinke}, and $\mathbf{e}$ is a unit vector in the direction of the polarization 
of the electric field of the photon. The final quantum state of AB is denoted ($n,j$), where $n$ and $j$ are the vibrational and
rotational quantum numbers, respectively.

\subsubsection{Quantum partial cross section and product state distribution}
\label{II.A.2}

Within the electric dipole approximation, the ($n,j$) state-resolved, or partial, absorption cross 
section is proportional to \cite{Schinke}
\\
\begin{equation}
\Sigma_E^{nj} = \left|\langle \bar{\Psi}_E^{nj} |\mathbf{d}.\mathbf{e}|\bar{\phi}_0 \rangle \right|^2.
\label{1}
\end{equation}
\\
$|\bar{\Psi}_E^{nj} \rangle$ is the state of inelastic scattering between AB and C at energy $E$, in the electronic excited state, and with outgoing free wave in channel ($n,j$). $|\bar{\phi}_0 \rangle$ is the rovibrational ground state in the electronic ground state. Final state populations are deduced from Eq.~\eqref{1} according to
\\
\begin{equation}
P_E^{nj} = \frac{\Sigma_E^{nj}}{\sum_{n,j}\;\Sigma_E^{nj}}.
\label{1a}
\end{equation}
\\
A very common approximation consists in replacing in Eq.~\eqref{1} $\mathbf{d}.\mathbf{e}$ by the modulous $d$ of $\mathbf{d}$, 
which amounts to suppose that $\mathbf{d}$ is parallel to $\mathbf{e}$ at the instant of the photon absorption. This approximation appears to be very satisfying provided than one is not interested in vector properties
\cite{Schinke,Alberto1,Alberto2,Guo2,Alberto3,Cong}. $\mathbf{d}$ is either parallel or perpendicular to the triatomic plane and $d$ does only depend on the configuration of ABC, i.e., $d \equiv d(R,r,\theta)$.

\subsubsection{Quantum partial cross section at time $t$}
\label{II.A.2}
 
Setting 
\begin{equation}
 \bar{\Phi}_0(\mathbf{R},\mathbf{r}) = d(R,r,\theta) \bar{\phi}_0(\mathbf{R},\mathbf{r}),
 \label{DD.2}
\end{equation}
Eq.~\eqref{1} reads
\begin{equation}
 \Sigma_E^{nj} = {\left| \int\; \mathbf{dR} \mathbf{dr}\;
 {\bar{\Psi}_E^{nj}}{}^* (\mathbf{R},\mathbf{r}) \bar{\Phi}_0(\mathbf{R},\mathbf{r}) \right|}^2.
 \label{DD.1}
\end{equation}
Writing the identity operator as
\begin{equation}
\hat{1} = e^{i\hat{\bar{H}}_e t/\hbar}e^{-i\hat{\bar{H}}_e t/\hbar}
\label{DD.3}
\end{equation}
\\
where the optical excitation defines the origin of time, and inserting this operator between the two states in Eq.~\eqref{DD.1} 
leads to
\\
\begin{equation}
\Sigma_E^{nj} = \left| \int\;\mathbf{dR} \mathbf{dr}\;
{\bar{\Psi}_E^{nj}}{}^*(\mathbf{R},\mathbf{r}) e^{iEt/\hbar}
\bar{\Phi}_t(\mathbf{R},\mathbf{r}) \right|^2.
\label{DD.4}
\end{equation}
\\
The phase factor $e^{iEt/\hbar}$ arises from the fact that $|\bar{\Psi}_E^{nj} \rangle$ is a stationary state of the system
in the electronic excited state. This factor is only written for clarity's sake, since its modulous is equal to one.  
$\bar{\Phi}_t(\mathbf{R},\mathbf{r})$ is the wave-packet obtained by propagating $\bar{\Phi}_0(\mathbf{R},\mathbf{r})$ 
in the electronic excited state during $t$.

It is shown in Appendix A that Eq.~\eqref{DD.4} can be rewritten as
\\
\begin{equation}
\Sigma_E^{nj} = \left| \int\;dR dr d\theta\;sin\theta\;{\Psi_E^{nj}}{}^*(R,r,\theta)\;\Phi_t(R,r,\theta) \right|^2
\label{2}
\end{equation}
\\
where $\Psi_E^{nj}(R,r,\theta)=8^{1/2}\pi Rr\bar{\Psi}_E^{nj}(\mathbf{R},\mathbf{r})$ and 
$\Phi_t(R,r,\theta)=8^{1/2}\pi Rr \bar{\Phi}_t(\mathbf{R},\mathbf{r})$. $\Psi_E^{nj}(R,r,\theta)$ and 
$\Phi_t(R,r,\theta)$ are shown in the same appendix to satisfy the Schr\"{o}dinger equations
\\
\begin{equation}
\hat{H}_e {\Psi}_E^{nj} = [\hat{T}+\hat{V}_e] {\Psi}_E^{nj} = E {\Psi}_E^{nj} 
\label{4}
\end{equation}
and
\begin{equation}
i\hbar \frac{d\Phi_t}{dt}=\hat{H}_e \Phi_t,
\label{DD.7}
\end{equation}
with
\begin{equation}
\hat{T} = - \frac{\hbar^2}{2\mu}\frac{\partial^2}{\partial R^2}
- \frac{\hbar^2}{2m}\frac{\partial^2}{\partial r^2}
- \frac{\hbar^2}{2Isin\theta}\frac{\partial}{\partial \theta}sin\theta\frac{\partial}{\partial \theta}.
\label{5}
\end{equation}
\\
$I$ is the reduced moment of inertia given by Eq.~\eqref{AA.10}.

$\Phi_t(R,r,\theta)$ is the wave-packet obtained by propagating $\Phi_0(R,r,\theta)$ during $t$.
Setting $\phi_0(R,r,\theta)=8^{1/2}\pi Rr \bar{\phi}_0(\mathbf{R},\mathbf{r})$, 
we deduce from Eq.~\eqref{DD.2} and Eq.~\eqref{AA.3a} at $t=0$, the identity
\\
\begin{equation}
\Phi_0(R,r,\theta) = d(R,r,\theta) \phi_0(R,r,\theta).
\label{5a}
\end{equation}
$\phi_0(R,r,\theta)$ is solution of
\begin{equation}
\hat{H}_g \phi_0 = [\hat{T}+\hat{V}_g] \phi_0 = E_0 \phi_0. 
\label{3}
\end{equation}
\\
Far in the asymptotic channel, the general expression of $\Psi_E^{nj}(R,r,\theta)$ is
\\
\begin{equation}
\Psi_E^{nj}(R,r,\theta) = \left[\frac{\mu}{2\pi\hbar^2 k_{nj}}\right]^{1/2}e^{ik_{nj}R}\chi_n(r)Y_{j}^0(\theta)
+\sum_{n'j'}\;S_{njn'j'}\;\left[\frac{\mu}{2\pi\hbar^2 k_{n'j'}}\right]^{1/2}e^{-ik_{n'j'}R}\chi_{n'}(r)Y_{j'}^0(\theta).
\label{6}
\end{equation}
\\
This expression assumes that the vibration and rotation motions of AB are uncoupled, a good approximation
provided that too highly excited rovibrational states are not available. $k_{nj}$ is defined by
\\
\begin{equation}
k_{nj}=\frac{1}{\hbar}[2\mu(E-E_{nj})]^{1/2}
\label{6a}
\end{equation}
\\
where $E_{nj}$ is the internal energy of AB in state $(n,j)$. 
%Its practical calculation is considered in section~\ref{V}. 
$\chi_n(r)$ is the $n$th excited vibrational state,
and $Y_{j}^0(\theta)$ is the $j$th spherical harmonic defined by
\\
\begin{equation}
Y_{j}^0(\theta)=\left[\frac{2j+1}{4\pi}\right]^{1/2}P_{j}(cos\theta).
\label{7}
\end{equation}
\\
$P_{j}(x)$ is the $j$th Legendre polynomial. The $S$-matrix element $S_{njn'j'}$ is the probability amplitude
to go from $(n',j')$ to $(n,j)$. The $\Psi_E^{nj}(R,r,\theta)$'s satisfy the usual orthogonality relations.

For $t$ tending to $+\infty$, $\Phi_t(R,r,\theta)$ entirely lies in the product channel and is moving outwards. 
The overlap in Eq.~\eqref{2} between the incoming part of $\Psi_E^{nj}(R,r,\theta)$ (see Eq.~\eqref{6}) and $\Phi_t(R,r,\theta)$ 
is thus zero. Hence, Eq.~\eqref{2} becomes
\\
\begin{equation}
\Sigma_E^{nj} =
\lim\limits_{t \to +\infty} \left| \int\;dR dr d\theta\;sin\theta\;\left[\frac{\mu}{2\pi\hbar^2 k_{nj}}\right]^{1/2}
e^{-ik_{nj}R}\chi_n(r)Y_{j}^0(\theta)\;
\Phi_t(R,r,\theta) \right|^2
\label{11}
\end{equation}
\\
($\chi_n(r)$ and $Y_{j}^0(\theta)$ being real functions, they are equal to their complex conjugate).

\subsubsection{Quantum partial cross section in terms of Wigner densities}
\label{II.A.3}

It is shown in Appendix B that a strictly equivalent phase space integral expression of Eq.~\eqref{11} is
\\
\begin{equation}
\Sigma_E^{nj} =
\lim\limits_{t \to +\infty} (2\pi \hbar)^3
\int\;\mathbf{d\Gamma}\; \rho_t(\mathbf{\Gamma})
\; \rho_{tr}(R,P)\; \rho_n(r,p)\; \rho_j(\theta,P_{\theta})
\label{12}
\end{equation}
\\
with $\mathbf{\Gamma}=(R,r,\theta,P,p,P_{\theta})$. 
$\rho_t(\mathbf{\Gamma})$ is the Wigner density related to $\Phi_t(R,r,\theta)$ through Eq.~\eqref{BB.3}.
$\rho_{tr}(R,P)$, $\rho_n(r,p)$ and $\rho_j(\theta,P_{\theta})$, respectively called translational, vibrational and rotational Wigner distributions, are given by 
\begin{equation}
\rho_{tr}(R,P) = \frac{1}{2\pi\hbar} \delta\Big[\frac{P^2}{2\mu}-\frac{\hbar^2 k_{nj}^2}{2\mu}\Big]\Theta(P),
\label{13}
\end{equation}
\begin{equation}
\rho_n(r,p) = \frac{1}{\pi \hbar}\int\;ds \; 
e^{2ips/\hbar}\;
\chi_n(r+s)\;
\chi_n(r-s)
\label{14}
\end{equation}
and
\begin{equation}
\rho_j(\theta,P_{\theta}) = \frac{1}{\pi \hbar}\int\;ds \; 
e^{2iP_{\theta}s/\hbar}\;
sin(\theta+s)\; Y_{j}^0(\theta+s)\;
sin(\theta-s)\; Y_{j}^0(\theta-s).
\label{15}
\end{equation}
\\
The present expression of $\rho_{tr}(R,P)$ (Eq.~\eqref{13}) appears to be different from the one in refs.~\cite{Heller2,Schinke2}, due
to different normalizations of translational states in Eq.~\eqref{6}.
$\rho_n(r,p)$ and $\rho_j(\theta,P_{\theta})$ are discussed in sections~\ref{II.B} and~\ref{II.C}.  
The argument of $Y_{j}^0$ being limited to the range [0,$\pi$], the constraints on $s$ in Eq.~\eqref{15} are 
$-\theta \le s \le \pi-\theta$ and $\theta-\pi \le s \le \theta$, or
\\
\begin{equation}
max[-\theta,\theta-\pi] \le s \le min[\theta,\pi-\theta].
\label{15a}
\end{equation}

\subsubsection{Passage to the semiclassical description}
\label{II.A.4}

We now introduce in the previous rigorous quantum formulation the following classical ingredient: 
we assume that the propagation of $\rho_{\tau}(\mathbf{\Gamma})$
from $\tau=0$ to $\tau=t$ is correctly ensured by classical mechanics, i.e., $\rho_{\tau}(\mathbf{\Gamma})$
reasonably satisfies Liouville equation \cite{Gold}. As is well known, this assumption is only valid over a short 
period of time \cite{Heller3,Schinke2,Truhlar} and its application should thus be limited to direct photodissociations. 
We shall come back to this important issue later below. 

In the framework of the previous assumption, we have according to Liouville theorem \cite{Gold}
\\
\begin{equation}
\rho_{t}(\mathbf{\Gamma})\mathbf{d\Gamma} = \rho_{0}(\mathbf{\Gamma_0})\mathbf{d\Gamma_0}.
\label{16}
\end{equation}
\\
In this identity, $\mathbf{\Gamma}$ should be understood as the dynamical state of ABC reached at time $t$ when starting from state
$\mathbf{\Gamma_0}$ at time 0. For both clarity and simplicity's sake, the components of $\mathbf{\Gamma}$ will be called 
$(R_t,r_t,\theta_t,P_t,p_t,{P_{\theta}}_t)$ in the rest of this section. In practice, they are determined by solving the Hamilton equations given in chapter 5 of Schinke's book (see Eqs. (5.3) and (5.4) where $m$, $\mu$, $V$ and
$\gamma$ are respectively denoted here by $\mu$, $m$, $V_e$ and $\theta$ ; these equations are clearly the classical analogs of 
Eqs.~\eqref{4} and \eqref{5} in the present work). Hamilton equations involve the classical Hamiltonian
\begin{equation}
H_e=\frac{P^2}{2\mu}+\frac{p^2}{2m}+\frac{{P_{\theta}}^2}{2I}+V_e(R,r,\theta)
\label{16a}
\end{equation}
\\
needed in the followings.

Eqs.~\eqref{12} and \eqref{16} finally lead to the semiclassical Wigner expression of Brown and Heller \cite{Heller2,Schinke2}
\\
\begin{equation}
\Sigma_E^{nj} =
\lim\limits_{t \to +\infty} (2\pi \hbar)^3
\int\;\mathbf{d\Gamma_0}\; \rho_0(\mathbf{\Gamma_0})
\; \rho_{tr}(R_t,P_t)\; \rho_n(r_t,p_t)\; \rho_j(\theta_t,{P_{\theta}}_t)
%\; \rho_{tr}[R(\mathbf{\Gamma_0},t),P(\mathbf{\Gamma_0},t)]\; 
%\rho_n[r(\mathbf{\Gamma_0},t),p(\mathbf{\Gamma_0},t)]\; 
%\rho_j[\theta(\mathbf{\Gamma_0},t),P_{\theta}(\mathbf{\Gamma_0},t)].
\label{17}
\end{equation}
\\
except that here, the rotation motion of AB is taken into account. 

A comment on $\theta_t$ is in order. As previously seen, $Y_{j}^0(\theta_t)$ is only defined for $\theta_t$ 
in the range [0,$\pi$], and we shall see later below (section~\ref{II.C}) that this is also the case of 
$\rho_j(\theta_t,{P_{\theta}}_t)$. 
However, $\theta_t$ is a (roughly) linear function of time in the asymptotic channel (see section~\ref{II.D}) and thus, it
eventually gets out of the previous interval. Replacing $\theta_t$ in Eq.~\eqref{17} by
\begin{equation}
\theta = \theta_t-\pi\;int(\theta_t/\pi)
\label{17a}
\end{equation}
if $int(\theta_t/\pi)$ is even, or    
\begin{equation}
\theta = \pi - [\theta_t-\pi\;int(\theta_t/\pi)]
\label{17b}
\end{equation}
if $int(\theta_t/\pi)$ is odd, keeps the ABC configuration unchanged and constraints $\theta$ to belong to the range [0,$\pi$].

Nevertheless, we shall see in section~\ref{II.D} that Eq.~\eqref{17}, despite its solid theoretical grounds and its elegance, 
cannot be applied as such. This is due to the degrading effect, consequence of the limitation of the validity of Eq.~\eqref{16} to
short times. Eq.~\eqref{17} needs a slight formal modification, proposed
after discussing to some extent the degrading effect.

\subsection{Vibrational Wigner densities}
\label{II.B}

Vibrational Wigner densities have been discussed elsewhere, in particular for the harmonic and Morse oscillators \cite{Stanek,Case,Schinke}. 
Therefore, we only concentrate on those involved in the process studied later in this work in order to check the validity of the semiclassical Wigner method, i.e., the reduced dimensionality model of methyl iodide photodissociation of Guo \cite{Guo2}.

\begin{figure}
 \includegraphics[width=180mm]{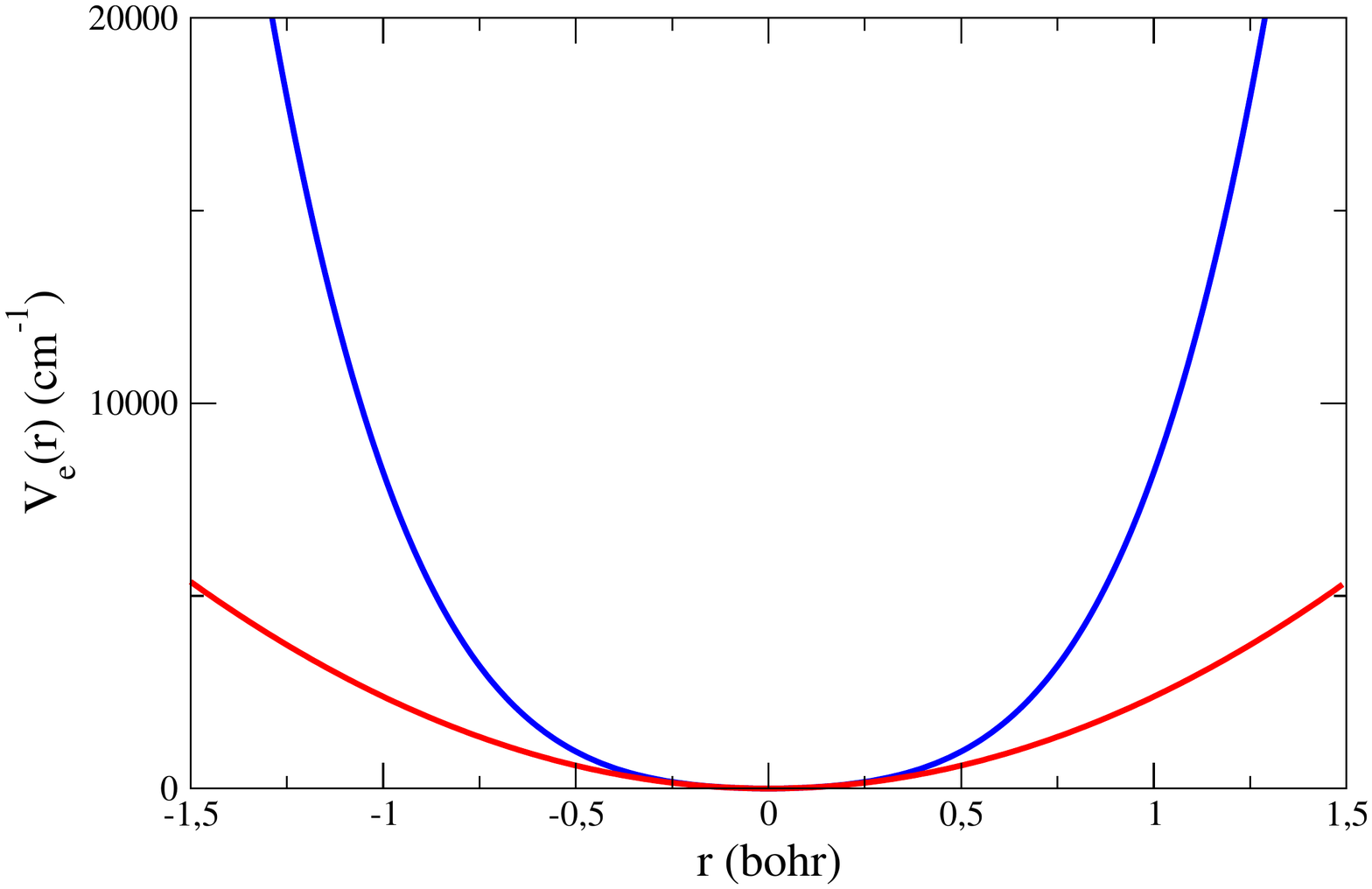}
 \caption{Blue curve: $r$-dependence, in the $^1Q_1$ electronic state, of the potential energy in the separated products 
 ($R \ge 12$ bohr) for the reduced dimensionality model of methyl iodide photodissociation of Guo \cite{Guo2}. 
 Its shape is the same in the $^3Q_0$ electronic state, with the difference that the bottom of the well is 
 7603.92 cm$^{-1}$ higher. Red curve: second order development of the previous potential.
 \label{figure1}}
\end{figure}

In this model, the methyl radical vibration motion is reduced to its umbrella mode, treated as if this was
the stretching mode of a pseudo diatomic molecule. The coordinate $r$ is between 
the centers-of-mass of the three hydrogen atoms and the carbon atom, assuming the $C_{3v}$ symmetry is preserved 
throughout the whole process (see section~\ref{V} for more details). The potential energy $v_e(r)$, represented in Fig.~\ref{figure1}
together with its second order development, 
clearly appears to be strongly anharmonic with a significant contribution from a fourth order term. 

The first vibrational states have been calculated by means of the Truhlar-Numerov algorithm \cite{Pablo}
and perspective views of the resulting Wigner distributions, estimated from Eq.~\eqref{14}
over a regular grid of 100 points, are displayed in Fig.~\ref{figure2} for the levels $n=0-5$.
The $r$ and $p$ axis are directed towards the right and left, respectively.
$r$ belongs to the range [-1.2,1.2] and $p_r$ to the range [-12,12], both in atomic units.
Despite the strong anharmonicity of $v_e(r)$, the shape similarity with the Wigner distributions of the 
harmonic oscillator is stricking \cite{Case,Schinke}. 

While for $n=0$, the Wigner density is always positive, it
takes negative values for $n\ge1$ in the crater-like areas. Craters are delimited by cones on the edge of which small summits are 
visible, particularly for $n\ge3$. Contrary to the harmonic case, contour levels do not exactly correspond to classical orbits, especially in the vicinity of the edge \cite{Schinke4}. This is at the origin of the degrading effect \cite{Truhlar,Schinke2}, 
as discussed further below. This effect will however appear to be much stronger in the case of the rotation motion.

\subsection{Rotational Wigner densities}
\label{II.C}

Perspective views of $\rho_j(\theta,P_{\theta})$, estimated from Eq.~\eqref{15} over a regular grid of 100 points, are shown in Fig.~\ref{figure3} for $j=0-3$, 6 and 10.
The $\theta$ and $P_{\theta}$ axis are oriented towards the right and left, respectively. 
$\theta$ belongs to the range [0,$\pi$], and $P_{\theta}$ to the range [-$15\hbar$,$15\hbar$].

At first sight, $\rho_j(\theta,P_{\theta})$ appears to involve $j$ narrow wells along
the symmetry axis of the distribution defined by $P_{\theta}=0$, separated by $j-1$ peaks 
(the second part of this statement is not true for $j=0$). These wells are in fact negative peaks
with magnitudes comparable with the ones of positive peaks. This is clearly seen in Fig.~\ref{figure4} where a front view
of $\rho_6(\theta,P_{\theta})$ along the $\theta$-direction is displayed. 
In this complementary view where the previous peaks are aligned, positive ones are in the light while negative ones
are in the shadow. We shall call \emph{central peaks} this chain of alternatively positive and negative spikes. 
In addition, $\rho_j(\theta,P_{\theta})$ for $j\ne0$ involves two positive \emph{lateral ridges} parallel to
the $\theta$-axis (note that for $j=0$, the two lateral ridges have merged with the central peak). 
As indicated in Fig.~\ref{figure4}, the summits of these ridges, located along the line $\theta=\pi/2$, 
appear to be roughly defined by $P_{\theta}\approx\pm \hbar[j(j+1)]^{1/2}$ in the upper and lower half plane,
respectively. This is actually true only for
large $j$'s, as illustrated in Fig.~\ref{figure4a} where the difference between (i) the exact
value of $P_{\theta}$ corresponding to the summit in the upper half plane 
and (ii) the quantum value $\hbar[j(j+1)]^{1/2}$ is seen to decrease to 0 in terms of $j$. 

For completeness, the semiclassical limit of the rotational Wigner density is considered in Appendix C.
The interest of this limit is that it clearly explains the main topological features of $\rho_j(\theta,P_{\theta})$
outlined above. Moreover, it provides analytical expressions which prove to be useful to make the link
between the semiclassical Wigner treatment and the method of Goursaud \emph{et al.} \cite{Goursaud}. The study of this link will
be published elsewhere \cite{Part5}.

%Moreover, we shall make use of it in order
%to establish later in this work the link between the 
%semiclassical Wigner treatment \cite{Heller2,Schinke2} 
%and the one of Goursaud \emph{et al.} \cite{Goursaud,Schinke}.

Due to the sign alternation of central peaks, their contribution to $\Sigma_E^{nj}$, given by Eq.~\eqref{17}, 
is expected to be negligible as compared to the one of lateral ridges provided that the distribution of 
the points ($\theta_t,{P_{\theta}}_t$) overlaps several peaks. In the semiclassical
limit of large $j$'s where the central peaks become very sharp (see $\rho_{10}(\theta,P_{\theta})$ in Fig.~\ref{figure3}), 
this will necessarily be the case. Consequently, one recovers the well known semiclassical rule stating that 
the values of $P_{\theta}$ close to $\pm \hbar[j(j+1)]^{1/2} \approx \pm \hbar(j+1/2)$ mostly contribute to the population of 
the $j^{th}$ rotational state. Conversely, the contribution of central peaks to $\Sigma_E^{nj}$ for $j=1-3$
cannot be excluded and the semiclassical rule fails (in the case where $j=0$, the semiclassical rule works for the
trivial reason that there is only one central massif).  

In the next two paragraphs where vibrational and rotational Wigner distributions are compared, 
we use the quantum number $j$ for both densities (and for the vibrational eigenstate as well). The
vibrational density $\rho_j(r,p)$ and
its rotational analog $\rho_j(\theta,P_{\theta})$ are then differentiated from their arguments only. 

\begin{figure}
 \includegraphics[width=200mm]{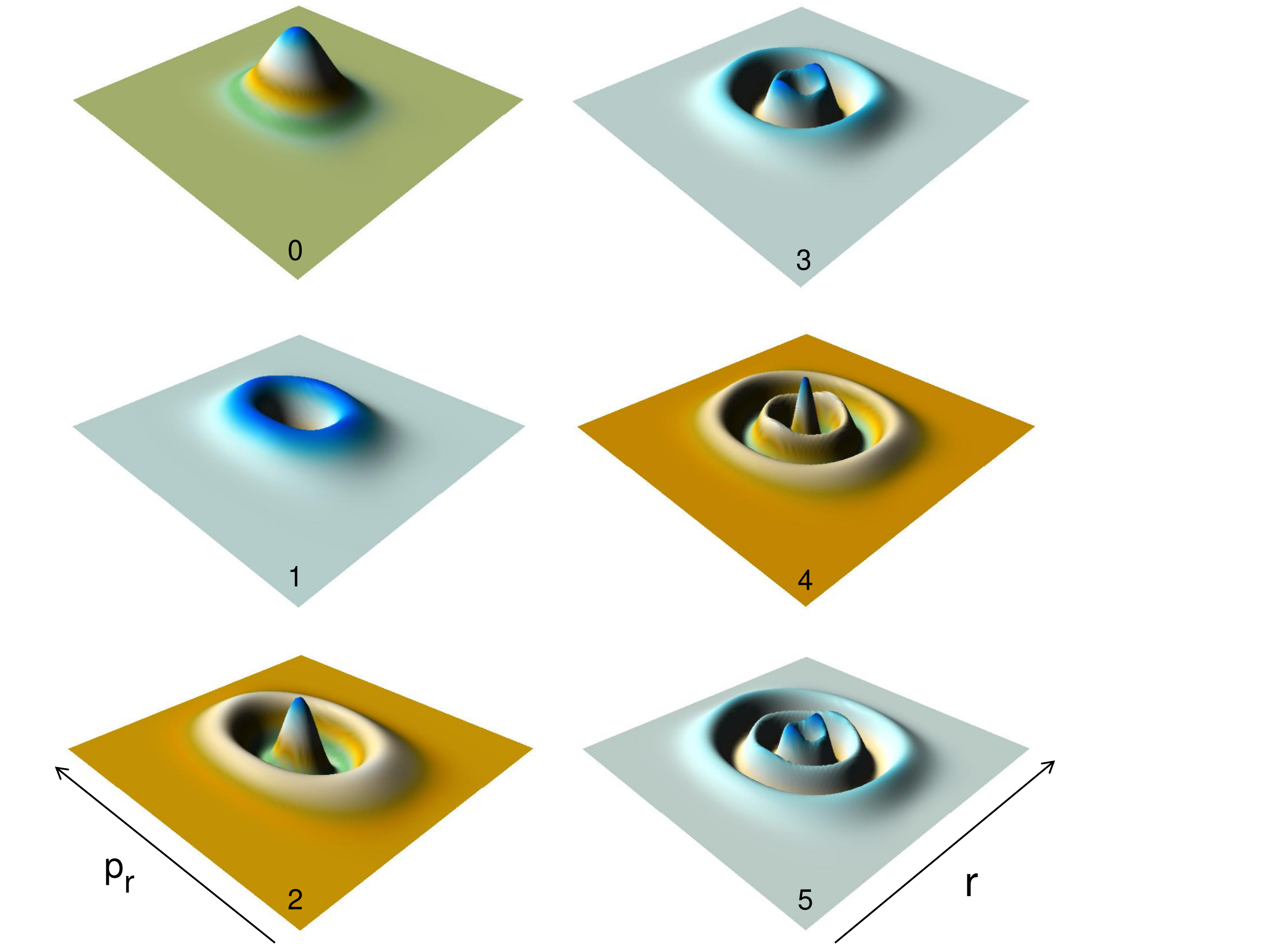}
 \caption{Perspective views of the vibrational Wigner distributions for the levels $n=0-5$ for the potential energy shown in Fig.~\ref{figure1} (blue curve). The $r$ and $p$ axis are directed towards the right and left, respectively.
  $r$ belongs to the range [-1.2,1.2], and $p_r$ to the range [-12,12], in atomic units.  
 \label{figure2}}
\end{figure}

\begin{figure}
 \includegraphics[width=200mm]{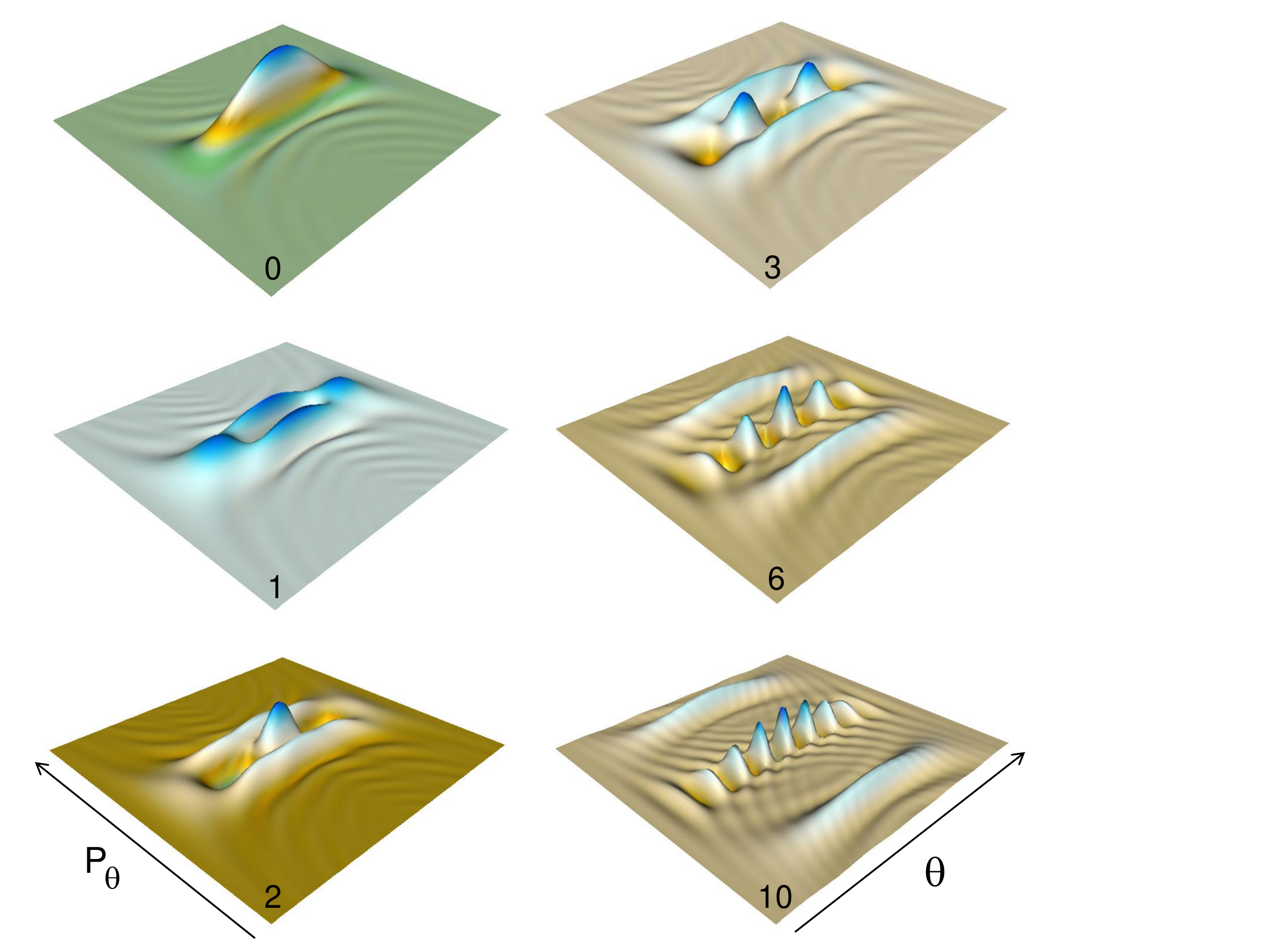}
 \caption{Perspective views of the rotational Wigner distributions for the states $j=0-3,6$ and $10$. 
 The $\theta$ and $P_{\theta}$ axis are oriented towards the right and left, respectively.  
 $\theta$ belongs to the range [0,$\pi$], and $P_{\theta}$ to the range [-$15\hbar$,$15\hbar$].
 \label{figure3}}
\end{figure}

\begin{figure}
 \includegraphics[width=180mm]{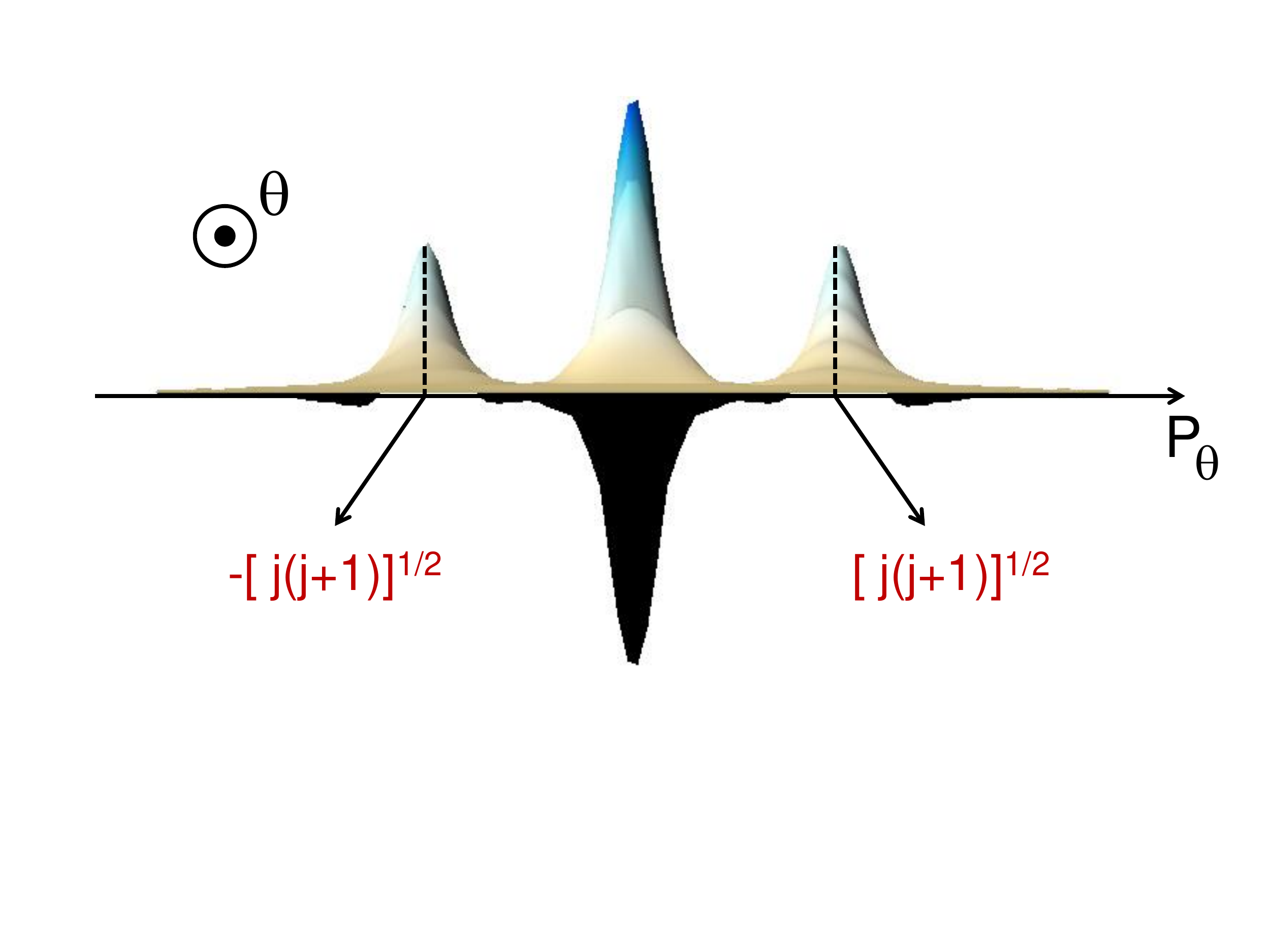}
 \caption{Front view of $\rho_6(\theta,P_{\theta})$ along the $\theta$-axis, directed towards us. 
The wells along the symmetry axis of the distribution correspond to negative peaks, not visible in Fig.~\ref{figure3}, 
having magnitudes comparable with those of the positive peaks. They form the black downward peak, resulting from
the alignment of 6 peaks. 
The two lateral peaks, forming ridges along the $\theta$-axis, 
have been found to be approximately centered at $\pm [j(j+1)]^{1/2}$ in $\hbar$ unit.  
 \label{figure4}}
\end{figure}

\begin{figure}
 \includegraphics[width=180mm]{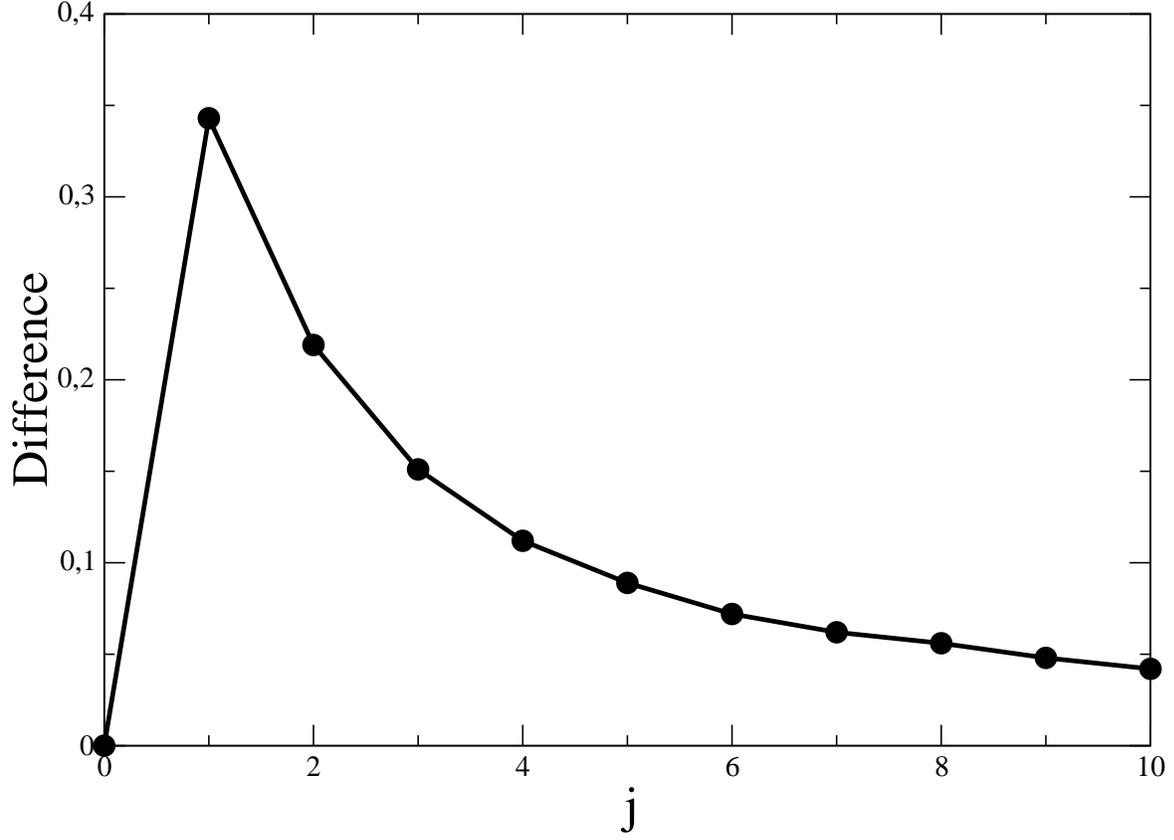}
 \caption{The difference between (i) the exact
value of $P_{\theta}$ corresponding to the summit of the rotational Wigner density in the upper half plane 
and (ii) the quantum value $\hbar[j(j+1)]^{1/2}$, tends to 0 as $j$ takes large values. In the particular case where $j=0$,
the lateral ridges have merged with the central peak, and their top is rigorously at the
expected quantum value, i.e., 0.   
 \label{figure4a}}
\end{figure}

The functions $sin\theta\; Y_{0}^0(\theta)$ and $sin\theta\; Y_{1}^0(\theta)$ are proportional to 
$sin\theta$ and $sin\theta cos\theta$, respectively. 
Topologically, these two functions are not very different from the ground and first excited vibrational states 
of a diatom. This is the reason why they result in densities
$\rho_{0}(\theta,P_{\theta})$ and $\rho_{1}(\theta,P_{\theta})$ 
bearing strong similarities with $\rho_{0}(r,p)$ and $\rho_{1}(r,p)$, respectively.
This is clearly seen when comparing Figs.~\ref{figure2} 
and~\ref{figure3}. For the ground states, both densities are bell-shaped and for the first
excited states, they both involve a crater with small summits lying on its edge (smaller for the vibrational
density than for the rotational one). 

On the other hand, 
$sin\theta\; Y_{j}^0(\theta)$ is significantly different from $\chi_j(r)$ for $j\ge2$, in particular
because of the $sin\theta$ term which makes the oscillation amplitude of $sin\theta\; Y_{j}^0(\theta)$ decrease when going away from 
$\pi/2$ (see Appendix C) while the one of $\chi_j(r)$ tends to increase when going away from the equilibrium geometry 
(at least up to the classical turning points). Consequently, $\rho_j(\theta,P_{\theta})$ appears to be very different 
from $\rho_j(r,p)$ for $j\ge2$, as can be observed when comparing Figs.~\ref{figure2} and~\ref{figure3}. 

For the free rotor, $P_{\theta}$ is a constant of motion and classical orbits are defined by straight lines parallel 
to the $\theta$-axis. It is thus clear from Figs.~\ref{figure3} and~\ref{figureC1} that the rotational Wigner density strongly
varies along a classical orbit, a fact responsible for the degrading effect, as shown in the next section. Note that the 
amplitude of the previous variation is much stronger than for the vibration motion. On average, classical orbits may indeed
be shown to be much closer to contour levels for the vibrational Wigner density than for the rotational one.   

The Wigner distribution of a rigid rotator has already been discussed in ref.~\cite{WigRR}. However, the context is 
quite different, hence leading to a different mathematical definition of the distribution. In addition, it seems that the notion of rotational Wigner density has never been introduced in the reaction dynamics field to date. Lastly, we note the shape 
similarity between the present rotational Wigner distributions and those for a symmetric infinite square well potential
(compare Fig. 1 in ref.~\cite{Lee} and Figs.~\ref{figure3} and~\ref{figureC1} in the present work).

\subsection{Franck-Condon process}
\label{II.D}

\subsubsection{System}
\label{II.D.1}

We now assume that $V_e$ is isotropic, i.e., does not depend on $\theta$. Moreover, we freeze the AB vibration
motion for simplicity's sake. AB is thus a rigid rotor the length of which is denoted $r_e$.
In addition, the reduced moment of inertia $I$, given by Eq.~\eqref{AA.10}, 
is supposed to reduce to $mr_e^2$. This is a good approximation, since generally, $R$ is already larger than $r_e$ at time 0, 
and unless C is much lighter than both A and B, $\mu$ is also larger than $m$. Lastly, the transition dipole moment is kept
at a constant value.

\subsubsection{Degrading effect}
\label{II.D.2}

Following the developments of 
section~\ref{II.A}, the $j$ state-resolved absorption cross section is proportional to
\\
\begin{equation}
\Sigma_E^j = \left| \int\;dR d\theta\;sin\theta\;{\Psi_E^j}^*(R,\theta)\;\Phi_0(R,\theta) \right|^2
\label{x1}
\end{equation}
\\
(see Eq.~\eqref{2} at time 0 without the $r$ coordinate). 
Due to the isotropy of $V_e$ which makes the radial and angular motions uncoupled, the scattering state $\Psi_E^j(R,\theta)$ 
can be written as
\\
\begin{equation}
\Psi_E^j(R,\theta) = U_E(R) Y_{j}^0(\theta).
\label{x2}
\end{equation}
Asymptotically,
\begin{equation}
U_E(R) = \left[\frac{\mu}{2\pi\hbar^2 k_j}\right]^{1/2}\left(e^{ik_jR}+e^{i\eta-ik_jR}\right)
\label{x3}
\end{equation}
with
\begin{equation}
k_{j}=\frac{1}{\hbar}\left[2\mu\left(E-\frac{\hbar^2j(j+1)}{2mr_e^2}\right)\right]^{1/2}.
\label{x4}
\end{equation}
\\
$\eta$ is the phase shift. 
We shall suppose that $V_e$ is sufficiently repulsive for the final translational energy to be much larger than the 
rotational energy $\hbar^2j(j+1)/(2mr_e^2)$. In other words, the product energy $E$ is mainly deposited into the translation
motion and the rotational energy is negligible as compared to $E$. Hence, $k_j$, and consequently $U_E(R)$, do not depend on $j$ in practice. Using Eq.~\eqref{x2} and following the developments of Appendix B, we can rewrite Eq.~\eqref{x1} as 
\\
\begin{equation}
\Sigma_E^{j} = (2\pi \hbar)^2
\int\;dR d\theta dP dP_{\theta}\; \rho_0(R,\theta,P,P_{\theta})
\; \rho_U(R,P)\; \rho_j(\theta,P_{\theta}).
\label{x5}
\end{equation}
\\
$\rho_0(R,\theta,P,P_{\theta})$ and $\rho_U(R,P)$ are the Wigner distributions associated with $\Phi_0(R,\theta)$
and $U_E(R)$ respectively, and $\rho_j(\theta,P_{\theta})$ has already been introduced. 

\begin{figure}
 \includegraphics[width=180mm]{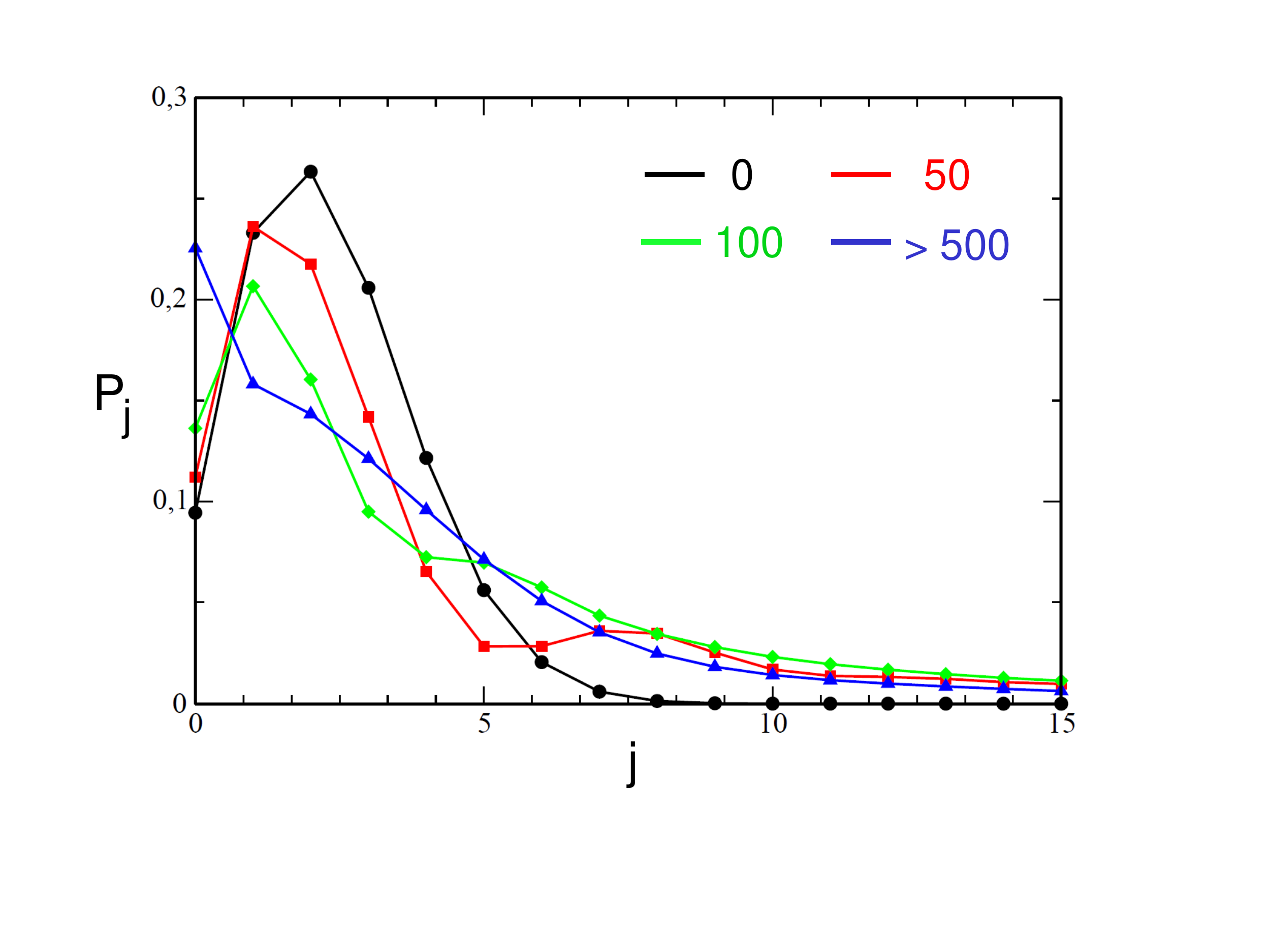}
 \caption{Time dependence of the rotational state distribution for the Franck-Condon process. Time is given in fs.   
 \label{figure5}}
\end{figure}

Within the harmonic approximation of $V_g$,
$\rho_0(R,\theta,P,P_{\theta})$ is found from Eq.~(5.20) of ref.~\cite{Schinke} 
to have the form
\\
\begin{equation}
\rho_0(R,\theta,P,P_{\theta}) = \frac{1}{(\pi \hbar)^2}
e^{-2\alpha_R(R-R_e)^2/\hbar}\;e^{-P^2/(2\alpha_R\hbar)^2}\;
e^{-2\alpha_{\theta}(\theta-\theta_e)^2/\hbar}\;e^{-P_{\theta}^2/(2\alpha_{\theta}\hbar)^2}.
\label{20}
\end{equation}
\\
Consequently, $\Sigma_E^{j}$ becomes
\begin{equation}
\Sigma_E^{j} \propto I_1 I_2
\label{20a}
\end{equation}
with
\begin{equation}
I_1=\int\;dR dP\;e^{-2\alpha_R(R-R_e)^2/\hbar}\;
e^{-P^2/(2\alpha_R\hbar)^2}\; \rho_U(R,P)
\label{20b}
\end{equation}
and
\begin{equation}
I_2=\int\;d\theta d{P_{\theta}}\;e^{-2\alpha_{\theta}(\theta-\theta_e)^2/\hbar}\;
e^{-{P_{\theta}}^2/(2\alpha_{\theta}\hbar)^2}\;\rho_j(\theta,{P_{\theta}}).
\label{20c}
\end{equation}
\\
Since $U_E(R)$, and consequently $\rho_U(R,P)$, do not depend on $j$, the $j$ dependence of $\Sigma_E^{j}$
is only due to $I_2$. Therefore, $I_1$ can be transfered into the proportionality factor of Eq.~\eqref{20a}, hence leading to  
\\
\begin{equation}
\Sigma_E^{j} \propto I_2.
\label{20d}
\end{equation}
\\
With $\alpha_{\theta} = 5$, corresponding to an initial angular distribution spreading over $\sim$ 20 degrees, and $\theta_e = 0$, we arrive at the distribution $P_j$ represented by the black curve in Fig.~\ref{figure5}. 
This distribution is ``exact'' within the assumptions of the present Franck-Condon model. 

Besides, $\Sigma_E^{j}$ is equally well given by
\\
\begin{equation}
\Sigma_E^j = \Big| \int\;dR d\theta\;sin\theta\;{\Psi_E^j}^*(R,\theta)\;\Phi_t(R,\theta) \Big|^2
\label{x7}
\end{equation}
\\
(see Eq.~\eqref{2}), which is Eq.~\eqref{x1} at time $t$ instead of time 0. 
Using Eq.~\eqref{x2} and following the developments of Appendix B, we can rewrite Eq.~\eqref{x7} as 
\\
\begin{equation}
\Sigma_E^j = (2\pi \hbar)^2
\int\;\mathbf{d\Gamma}\; \rho_t(\mathbf{\Gamma})
\; \rho_U(R,P)\; \rho_j(\theta,P_{\theta})
\label{x8}
\end{equation}
\\
where $\mathbf{\Gamma}=(R,\theta,P,P_{\theta})$. Just as Eqs.~\eqref{12} and \eqref{16} lead to Eq.~\eqref{17}, Eqs.~\eqref{16} and~\eqref{x8} lead to
\\
\begin{equation}
\Sigma_E^{j} = (2\pi \hbar)^2
\int\;\mathbf{d\Gamma_0}\; \rho_0(\mathbf{\Gamma_0})
\; \rho_U(R_t,P_t)\;\rho_j(\theta_t,{P_{\theta}}_t),
\label{x9}
\end{equation}
\\
$\mathbf{\Gamma_0}$ being the value of $\mathbf{\Gamma}=(R,\theta,P,P_{\theta})$ at time 0. When making
$t$ tend to infinity, one recovers the semiclassical Wigner expression analogous to Eq.~\eqref{17} in the present
case where the vibration motion is frozen. 

Since the radial motion is uncoupled with the angular motion, $R_t$ and $P_t$ are functions of $R_0$, $P_0$ and $t$,
while $\theta_t$ and ${P_{\theta}}_t$ are functions of $\theta_0$, ${P_{\theta}}_0$ and $t$. 
Using Eq.~\eqref{20}, Eq.~\eqref{x9} can thus be rewritten as
\\
\begin{equation}
\Sigma_E^{j} \propto I_1(t)I_2(t)
\label{27}
\end{equation}
with
\begin{equation}
I_1(t)=\int\;dR_0dP_0\;e^{-2\alpha_R(R_0-R_e)^2/\hbar}\;
e^{-P_0^2/(2\alpha_R\hbar)^2}\; \rho_U(R_t,P_t)
\label{28}
\end{equation}
and
\begin{equation}
I_2(t)=\int\;d\theta_0 d{P_{\theta}}_0\;e^{-2\alpha_{\theta}(\theta_0-\theta_e)^2/\hbar}\;
e^{-{P_{\theta}}_0^2/(2\alpha_{\theta}\hbar)^2}\;\rho_j(\theta_t,{P_{\theta}}_t).
\label{29}
\end{equation}
\\
Since $\rho_U(R_t,P_t)$ does not depend on $j$, the $j$ dependence of $\Sigma_E^{j}$
is only due to $I_2(t)$. Therefore, $I_1(t)$ can be transfered into the proportionality factor of Eq.~\eqref{27}, leading thereby to  
\\
\begin{equation}
\Sigma_E^{j} \propto I_2(t).
\label{29a}
\end{equation}
\\
AB rotating freely after the photon absorption, $\theta_t$ and ${P_{\theta}}_t$ are given by
\\
\begin{equation}
\theta_t = \theta_0+\frac{{P_{\theta}}_0}{mr_e^2}t
\label{25}
\end{equation}
and
\begin{equation}
{P_{\theta}}_t = {P_{\theta}}_0.
\label{26}
\end{equation}
\\
Eq.~\eqref{25} is the solution of Eq.~(5.4c) of ref.~\cite{Schinke}, remembering that the reduced moment of inertia $I$ 
(see Eq.~\eqref{AA.10}) reduces here to $mr_e^2$, as stated at the beginning of this section. Therefore, we finally arrive at
\\
\begin{equation}
\Sigma_E^{j} \propto
\int\;d\theta_0 d{P_{\theta}}_0\;e^{-2\alpha_{\theta}(\theta_0-\theta_e)^2/\hbar}\;
e^{-{P_{\theta}}_0^2/(2\alpha_{\theta}\hbar)^2}\;\rho_j\Big[\theta_0+\frac{{P_{\theta}}_0}{mr_e^2}t,{P_{\theta}}_0\Big].
\label{30a}
\end{equation}
\\
Note that Eqs.~\eqref{20c} and~\eqref{20d} are recovered from Eq.~\eqref{30a} at time 0. 

With the hypothetical values $m=5$ g.mol$^{-1}$ and $r_e=1$ $\AA$, Eq.~\eqref{30a} applied at $t=50,100$ and 500 fs
leads to the rotational state distributions displayed in Fig.~\ref{figure5}, in addition to the ``exact'' one at time 0. 
Beyond 500 fs, the distribution does not evolve.
The degrading effect previously outlined is patent, the distribution getting strongly altered after only 50 fs. 
This is a clear illustration of the inability of Eq.~\eqref{17}, as such, to correctly describe partial cross sections.

We note from Eq.~\eqref{30a} that the degrading effect is due to the strong variation of the rotational Wigner distribution 
along classical orbits, as previously outlined. The analogous variation being much weaker for the vibration motion, so is the corresponding degrading effect.

\subsubsection{Using Brown and Heller expression at a large distance rather than at a large time}
\label{II.D.3}

For the previous Franck-Condon process, it is equivalent to state that the distribution is exact at time 0, or at the time 
beyond which $P_{\theta}$ ceases to vary. For a general process, the latter corresponds to the instant where the 
system crosses the frontier separating the interaction region from the free products. This frontier being well
defined by a given value $R_f$ of $R$, we shall use instead of Brown and Heller expression~\eqref{17}, 
\\
\begin{equation}
\Sigma_E^{nj} = (2\pi \hbar)^3
\int\;\mathbf{d\Gamma_0}\; \rho_0(\mathbf{\Gamma_0})
\; \rho_{tr}(R_f,P_f)\; \rho_n(r_f,p_f)\; \rho_j(\theta_f,{P_{\theta}}_f),
\label{31}
\end{equation}
\\
where $P_f$ is the value of $P$ at $R_f$ when starting 
from $\mathbf{\Gamma_0}$,  with a similar definition for $r_f$, $p_f$, $\theta_f$ and ${P_{\theta}}_f$.
One will not forget to substitute $\theta$ for $\theta_f$ according to Eqs.~\eqref{17a} and~\eqref{17b}
(with $\theta_f$ instead of $\theta_t$).   
The above expression is assumed to minimize the degrading effect, and is exact in the Franck-Condon limit. 

In practice, the delta function in $\rho_{tr}(R_f,P_f)$ (see Eq.~\eqref{13}) may be replaced by a bin
centered at $\hbar^2 k_{nj}^2/(2\mu)$, much narrower than the distribution of $P_f^2/(2\mu)$. In this work, the bin width
is taken at five percent of the full-width-at-half-maximum (FWHM) of the previous distribution.

As stated in the introduction, we call the present method forward I and apply it to the photodissociation of methyl iodide in section~\ref{V}.

\subsection{Approximate formulation}
\label{II.E}

Eq.~\eqref{31} contains a delta distribution through the translational Wigner function
$\rho_{tr}(R_f,P_f)$, given by Eq.~\eqref{13}. This term, replaced by a narrow Gaussian
or a thin box in practical calculations \cite{Heller2,Schinke2}, makes them heavy. However, at the exit 
of the interaction region, the classical Hamiltonian~\eqref{16a} reads
\\
\begin{equation}
H_e=\frac{P^2}{2\mu}+E_{int}(r,p,P_{\theta})
\label{32}
\end{equation}
with
\begin{equation}
E_{int}(r,p,P_{\theta})=\frac{p^2}{2m}+v_e(r)+\frac{{P_{\theta}}^2}{2mr^2}.
\label{33}
\end{equation}
\\
$E_{int}(r,p,P_{\theta})$ represents the internal energy of AB. From Eqs.~\eqref{6a} and~\eqref{32}, 
one may thus rewrite $\rho_{tr}(R_f,P_f)$ as
\\
\begin{equation}
\rho_{tr}(R_f,P_f) = 
\frac{1}{2\pi\hbar} \delta\Big[H_e-E+E_{nj}-E_{int}(r_f,p_f,{P_{\theta}}_f)\Big].
\label{34}
\end{equation}
\\
Note the disappearence of the term $\Theta(P_f)$ as compared to Eq.~\eqref{13}, for $P_f$ is necessarily positive. 

For not too excited rovibrational states, $E_{nj}$ is very well approximated by the sum of $E^v_n$, the vibrational
energy corresponding to the $n^{th}$ state of the non rotating AB diatom, and the rotational energy $E^r_j=\hbar^2j(j+1)/(2mr_e^2)$. 
The values of $r_f$ and $p_f$ corresponding to $E^v_n$ define an elliptic-like curve in the $(r,p)$ plane 
(a true ellipse for a purely harmonic oscillator) while
the values of ${P_{\theta}}_f$ corresponding to $E^r_j$ are $\sim \pm \hbar (j+1/2)$.
However, $r_f$, $p_f$ and ${P_{\theta}}_f$ are weighted in Eq.~\eqref{31} by  $\rho_n(r_f,p_f)$ and $\rho_j(\theta_f,{P_{\theta}}_f)$
which broadly extend around the previous values. Consequently, one expects $E_{int}(r_f,p_f,{P_{\theta}}_f)$ 
to have roughly the same chance to be larger or lower than $E_{nj}$. In other words, their average difference should be negligible
as compared to $E$, so the partial cross sections obtained by means of Eq.~\eqref{34} or 
\\
\begin{equation}
\rho_{tr}(R_f,P_f) = 
\frac{1}{2\pi\hbar} \delta(H_e-E)
\label{35}
\end{equation}
\\
should not be very different.

The interest of this approximation is that $H_e$, as a constant of motion, can be expressed in terms of $\mathbf{\Gamma_0}$. 
One can use this fact to analytically integrate with respect to $P_0$ and $p_0$ as follows. Setting 
\\
\begin{equation}
P_0 = \mu^{1/2} \eta cos\beta
\label{36}
\end{equation}
and
\begin{equation}
p_0 = m^{1/2} \eta sin\beta
\label{37}
\end{equation}
\\
with $\beta$ in the range [0,$2\pi$], Eq.~\eqref{16a} becomes
\\
\begin{equation}
H_e=\frac{\eta^2}{2}+\frac{{P_{\theta}}_0^2}{2I}+V_e(R_0,r_0,\theta_0).
\label{38}
\end{equation}
\\
From Eqs.~\eqref{31} and \eqref{35}-\eqref{38}, the partial cross section $\Sigma_E^{nj}$ reads
\\
\begin{equation}
\Sigma_E^{nj} \propto 
\int\;\eta d\eta d\beta dR_0 dr_0 d\theta_0 d{P_{\theta}}_0\; \rho_0(\mathbf{\Gamma_0})
\; \delta\Big[\frac{\eta^2}{2}-Q\Big]\; \rho_n(r_f,p_f)\; \rho_j(\theta_f,{P_{\theta}}_f)
\label{39}
\end{equation}
with
\begin{equation}
Q=E-\frac{{P_{\theta}}_0^2}{2I}-V_e(R_0,r_0,\theta_0).
\label{40}
\end{equation}
\\
Replacing $\eta d\eta$ in Eq.~\eqref{39} by $d\eta^2/2$ makes the delta function disappear and we
finally arrive at the useful expression
\\
\begin{equation}
\Sigma_E^{nj} \propto 
\int\;d\beta dR_0 dr_0 d\theta_0 d{P_{\theta}}_0\; \rho_0(\mathbf{\Gamma_0})
\; \rho_n(r_f,p_f)\; \rho_j(\theta_f,{P_{\theta}}_f).
\label{41}
\end{equation}
\\
The values of $R_0$, $r_0$, $\theta_0$ and ${P_{\theta}}_0$ contributing to the integral are
those making $Q$ positive or zero. $P_0$ and $p_0$, which complete the set of initial conditions, are determined by means of Eqs.~\eqref{36} and \eqref{37} with 
\\
\begin{equation}
\eta=(2Q)^{1/2}.
\label{42}
\end{equation}
\\
We call the present method forward II. 

Another possibility would have been to follow Goursaud \emph{et al.} \cite{Goursaud} and integrate over one of the two momenta
$P_0$ or $p_0$. But a term diverging at the boundaries of the available phase space volume would have appear in the integrand,
rendering thereby the numerical calculation of $\Sigma_E^{nj}$ more tricky. 

%It is interesting to consider the classical limit of the forward II method by substituting in 
%\\
%\begin{equation}
%\Sigma_E^{nj} = 
%\int\;\mathbf{d\Gamma_0}\; \rho_0(\mathbf{\Gamma_0})
%\; \delta(H_e-E)\; \rho_n(r_f,p_f)\; \rho_j(\theta_f,{P_{\theta}}_f),
%\label{42a}
%\end{equation}
%\\
%obtained from Eqs.~\eqref{31} and \eqref{35}, semiclassical approximations
%of $\rho_n(r_f,p_f)$ and $\rho_j(\theta_f,{P_{\theta}}_f)$. The first one is 
%\\
%\begin{equation}
%\rho_n(r_f,p_f) \propto \delta(n(r_f,p_f)-n),
%\label{42b}
%\end{equation}
%\\
%derived by Berry \cite{Berry}, valid for large values of $n$. The second one is approximated by the product 
%\\
%\begin{equation}
%\rho_j(\theta,P_{\theta}) \propto {sin\theta}^{5/2}(sinc[\pi (|P_{\theta}|-j)]+sinc[\pi (|P_{\theta}|-j-1)]).
%\label{42c}
%\end{equation}
%\\

\section{Backward semiclassical Wigner approach}
\label{III}

From Eqs.~\eqref{31} and \eqref{34}, we have
\\
\begin{equation}
\Sigma_E^{nj} = (2\pi \hbar)^3
\int\;\mathbf{d\Gamma_0}\; \rho_0(\mathbf{\Gamma_0})
\; \delta\Big[H_e-E+E_{nj}-E_{int}(r_f,p_f,{P_{\theta}}_f)\Big]\; \rho_n(r_f,p_f)\; \rho_j(\theta_f,{P_{\theta}}_f).
\label{43}
\end{equation}
\\
In refs.~\cite{Part1,Part2,Part3,Gutz,Kay1,Kay2}, it is shown that an alternative set of coordinates to $\mathbf{\Gamma_0}$ is
$(t,H_e,r_f,p_f,\theta_f,{P_{\theta}}_f)$. The origin of time corresponds to the instant where the system is at $R_f$.
The quadruplet $(r_f,p_f,\theta_f,{P_{\theta}}_f)$ specifies the internal state of AB at time 0 and $H_e$ forces 
$P_f$ to take the value 
\\
\begin{equation}
P_f=[2\mu(H_e-E_{int}(r_f,p_f,{P_{\theta}}_f))]^{1/2}
\label{44}
\end{equation}
\\
(see Eq.~\eqref{32}). $\mathbf{\Gamma}_f=(R_f,P_f,r_f,p_f,\theta_f,{P_{\theta}}_f)$ lies along a given trajectory.
Now, any point along this trajectory can be reached from $\mathbf{\Gamma}_f$ by moving along the trajectory  
a given period of time $|t|$ either forward ($t > 0$) or backward ($t < 0$). In other words, for a given $R_f$,
$(H_e,r_f,p_f,\theta_f,{P_{\theta}}_f)$ imposes the classical path, and $t$ the location along it. 
Consequently, $(t,H_e,r_f,p_f,\theta_f,{P_{\theta}}_f)$ allows to span the whole phase space. 

In addition to that, one may show the important property \cite{Part1,Part2,Part3,Gutz,Kay1,Kay2}
\\
\begin{equation}
\mathbf{d\Gamma_0}=dt dH_e dr_f dp_f d\theta_f d{P_{\theta}}_f.
\label{45}
\end{equation}
\\
The exact demonstration of the above identity
is not given in the previous references, but it closely follows, for example, the developments in Appendix C of ref.~\cite{Part3}
for different (though partly related) coordinates. From Eqs.~\eqref{43} and \eqref{45} and the straightforward integration 
over $H_e$, we finally arrive at
\\
\begin{equation}
\Sigma_E^{nj} = (2\pi \hbar)^3
\int\;dr_f dp_f d\theta_f d{P_{\theta}}_f 
\; \rho_n(r_f,p_f)\; \rho_j(\theta_f,{P_{\theta}}_f)\int_{-\infty}^0\;dt\; \rho_0(\mathbf{\Gamma}_t).
\label{46}
\end{equation}
\\
Integration over $H_e$ forces the latter to be equal to $E-E_{nj}+E_{int}(r_f,p_f,{P_{\theta}}_f)$. 
From Eqs.~\eqref{6a} and~\eqref{44}, we thus have  
\\
\begin{equation}
P_f=\hbar k_{nj}.
\label{47}
\end{equation}
\\
%which in addition to the four first variables of integration 
%and the fact that trajectories are started at $R_f$,
%achieves to specify their initial conditions $\mathbf{\Gamma}_f$. 
At last, $\mathbf{\Gamma}_t$ in Eq.~\eqref{46} is the value of $\mathbf{\Gamma}$ at time $t$ when starting from $\mathbf{\Gamma}_f$ at time 0 (the meaning of $\mathbf{\Gamma}_t$ is thus different here and in section~\ref{II.A}).

To summarize, the internal state $(r_f,p_f,\theta_f,{P_{\theta}}_f)$ of AB is randomly chosen within appropriate boundaries.
Together with Eq.~\eqref{47}, they allow to generate a trajectory from $R_f$ at time 0. The trajectory is then propagated
backward in time, i.e., in the direction of the reagent molecule, and $\rho_0(\mathbf{\Gamma}_t)$ 
is time-integrated until the trajectory 
recrosses $R_f$ towards the products. 
The result is multiplied by the statistical weight $\rho_n(r_f,p_f)\; \rho_j(\theta_f,{P_{\theta}}_f)$
in order to get the integrand of Eq.~\eqref{46}. 
A simple Monte-Carlo procedure can then be used to estimate $\Sigma_E^{nj}$.  

In practice, the power of the backward approach is limited by the fact that one
cannot a priori guess which values of $r_f$, $p_f$, $\theta_f$ and ${P_{\theta}}_f$ 
lead to trajectories crossing the Wigner region,
corresponding to the phase space volume where Wigner distribution $\rho_0(\mathbf{\Gamma})$ takes significant values
(there is some arbitrariness in this definition). Therefore, a straightforward application of
Eq.~\eqref{46} may require running a large amount of useless trajectories that do not contribute to $\Sigma_E^{nj}$. To go round this difficulty, one may first apply the forward approach in order to determine the boundaries of $r_f$, $p_f$, $\theta_f$ and ${P_{\theta}}_f$ contributing to $\Sigma_E^{nj}$, and then apply the backward method with these variables selected within the previous boundaries. 

The practical method used here to perform this selection is as follows. First, one runs a few thousand trajectories, say $N$, within the forward I method, leading to the same number of final points ($r_k$, $p_k$, $\theta_k$, ${P_{\theta}}_k$), $k=1,...,N$,
at $R=R_f$. Next, one randomly generates a point ($r_f$, $p_f$, $\theta_f$, ${P_{\theta}}_f$) and checks whether it lies
within at least one of the small rectangular cuboids defined by $|r_f-r_k| \le \eta_r$, $|p_f-p_k| \le \eta_p$,
$|\theta_f-\theta_k| \le \eta_\theta$, $|{P_{\theta}}_f-{P_{\theta}}_k| \le \eta_{P_{\theta}}$, $k=1,...,N$. If so, this 
point serves as initial conditions together with $R=R_f$ and Eq.~\eqref{47}. Otherwise, one randomly generates another 
point and so on. The parameters $\eta_r$, $\eta_p$, $\eta_\theta$ and $\eta_{P_{\theta}}$ have to be chosen from
a visual inspection of the domain covered by the rectangular projections of the cuboids in the planes ($r,p$) and ($\theta,P_{\theta}$),
separately. The parameters must be large enough for the domains to be compact, as they appear to be when running millions of trajectories within the forward I method. 

Finally, we have found that for the model of methyl iodide photodissociation considered in section~\ref{V}, 
the values of $k_{nj}$ appear to be almost independent on $j$ (see section~\ref{II.D} for an explanation),
a bit less on $n$. In such a case, a single batch of trajectory can be run with $P_f=k_{n0}$ for calculating all the 
$\Sigma_E^{nj}$s corresponding to $n$.

\section{The standard method of Goursaud \emph{et al.}}
\label{IV}

The method of Goursaud \emph{et al.} was initially applied to a bi-dimensional model of triatomic ion fragmentation
with frozen valence angle. The method was later extended by
Schinke to realistic three-dimensional triatomic photodissociations \cite{Schinkesintheta}.
Within this approach, which is detailed in Chapter 5 of Schinke's book \cite{Schinke}, $\Sigma_E^{nj}$ is given by
\\
\begin{equation}
\Sigma_E^{nj} = (1+\delta_{j0})\;
\int\;\mathbf{d\Gamma_0}\; \rho_0(\mathbf{\Gamma_0})\; sin\theta_0
\; \delta(H_e-E)\; \delta(n(r_f,p_f)-n)\; \delta(|{P_{\theta}}_f|-j)
\label{49}
\end{equation}
\\
where the only quantity not defined until now is the final vibrational action $n(r_f,p_f)$ of AB, given by
\\
\begin{equation}
n(r_f,p_f) = \frac{2}{h}\int_{r_{in}}^{r_{out}}\;dr\;p(r)-\frac{1}{2}
\label{50}
\end{equation}
with
\begin{equation}
p(r) = \Big(2 m \Big[E_{int}(r_f,p_f,{P_{\theta}}_f)-\frac{{P_{\theta}}_f^2}{2 m r^2}-v_e(r)\Big]\Big)^{1/2}.
\label{51}
\end{equation}
\\ 
$r_{in}$ and $r_{out}$ are the values of $r$ at the inner and outer turning points. The internal energy 
$E_{int}(r_f,p_f,{P_{\theta}}_f)$ is given by Eq.~\eqref{33}.

In practice, the delta functions are replaced by standard bins (SB) of unit height and width, or
Gaussians the FWHM of which is usually taken at 10$\%$. The second procedure is called Gaussian
binning (GB) \cite{GB1,GB2,GB3,GB4,GB5,GB6,GB7}. These procedures are discussed at length in ref.~\cite{Part3}.
Since no Wigner distributions are used to weight the final dynamical states, the present method is 
more classical than the previous ones.   

The transition moment being absorbed in $\rho_0(\mathbf{\Gamma_0})$, Eq.~\eqref{49} is similar to Eq.~(5.23) 
in ref.~\cite{Schinke} (see also Eq.~(5.22)). The main difference is that the degeneracy factor $(1+\delta_{j0})$
has been added in the present work. Note that Schinke and co-workers also include this factor in practice \cite{Schinke3}.

The degeneracy factor doubles the integral in Eq.~\eqref{49} for $j=0$. This counterbalances the fact that due to the 
$\delta(|{P_{\theta}}_f|-j)$ term,
two values of ${P_{\theta}}_f$ contribute to the integral for $j>0$ ($\pm j$), against only one for $j=0$. 
The interest of this factor clearly appears in the purely statistical limit where $n(r_f,p_f)$ and ${P_{\theta}}_f$ 
are random variables. $\Sigma_E^{nj}$ being proportional
to the density of probability to get $n(r_f,p_f)=n$ and $|{P_{\theta}}_f|=j$ (see Eq.~\eqref{49}), all the $\Sigma_E^{nj}$'s are
equal (without the degeneracy factor, $\Sigma_E^{n0}$ would be half $\Sigma_E^{nj\ne0}$), a result in conformity with the 
quantum phase space theory expectations of equal final state populations \cite{Tomas,Pola1}.

\section{Photodissociation of methyl iodide}
\label{V}

We briefly summarize the main features of the reduced-dimensionality model of methyl iodide photodissociation
and the quantum dynamical method used to check the validity of the semiclassical Wigner method. More details 
can be found in the references mentioned below.

\subsection{Model}
\label{V.A}

The CH$_3$I molecule is considered as a CXI pseudotriatomic molecule \cite{ru09,gu91,Guo2}, 
the pseudoatom X=H$_3$ being located at the center-of-mass (CM) of   
the three H atoms. $\mathbf{R}$ is the vector between the CH$_3$
(or C$-$X) CM and I and $\mathbf{r}$ is the one between X and C. $r$ represents the
umbrella bend of the C$-$H$_3$ group. 
%The angle $\theta$ between $\mathbf{R}$ and $\mathbf{r}$ 
%accounts for the X$-$C$-$I bending motion. 

Photodissociation of CH$_3$I is assumed to take place upon 
optical excitation at $266$ nm (A band) from the $\tilde{X}^1A_1$ ground 
electronic state to the $^3Q_0$, $^1Q_1$ and $^3Q_1$ excited 
electronic states. Taking however into account that absorption to the 
$^3Q_1$ state is relatively small at $266$ nm, the present simulations 
only involve the $^3Q_0$, and $^1Q_1$ excited electronic states, in addition
to the ground state $\tilde{X}^1A_1$.
The $\tilde{X}^1A_1$ and $^1Q_1$ electronic surfaces correlate  
asymptotically with the CH$_3$ + I($^2P_{3/2}$) products, while 
the $^3Q_0$ surface correlates with the CH$_3$ + I*($^2P_{1/2}$) 
products. In addition to the coupling of $\tilde{X}^1A_1$ to the   
excited electronic states through electric-dipole moments,  
the $^3Q_0$ and $^1Q_1$ states are non adiabatically coupled. 
Taking into account the transitions between these
states in the semiclassical Wigner method would require treating these 
by means of a semiclassical approach of non adiabatic transitions
such as, for instance, the Landau-Zener model \cite{Nik}, the Zhu-Nakamura model \cite{Nak},
or the surface hopping method of Tully \cite{Tul,Bar}. For clarity's sake, however,
we artificially take at zero the coupling between the $^3Q_0$ and $^1Q_1$ states,
focusing our attention on the semiclassical Wigner method 
for fragmentations taking place on a single excited electronic state.
However, we plan to extend this method to processes
involving non adiabatic transitions in a near future. 

Upon optical excitation at $266$ nm, the energies available to the final products
in the $^3Q_0$ and $^1Q_1$ states are 11258.53 cm-1 and 18862.09 cm-1, respectively.

High-quality {\it ab initio} calculations have been used to 
model the three electronic potential energy surfaces (PESs) involved 
in the calculations. In the case of the $\tilde{X}^1A_1$ ground 
state, the PES ($V_g(R,r,\theta)$) is represented as a sum of three
potential interactions in the $R_{C-I}$ (the C$-$I nuclear
distance), $r$ and $\theta$ coordinates, respectively. The
interaction potential in the $R_{C-I}$ coordinate is taken from
the recently reported 2D ground-state potential for CH$_3$I,
obtained by means of multireference spin-orbit configuration
interaction {\it ab initio} calculations \cite{al07I,alpri}. The
potential interactions in the $r$ and $\theta$ coordinates
are represented by harmonic oscillator functions \cite{Guo2,Alberto1}. 

The PESs for the excited electronic states $^3Q_0$ and $^1Q_1$ 
($V_e(R,r,\theta)$), are the 
{\it ab initio} PES constructed by Xie et al. \cite{xi00}. These  
are improved versions of the previous nine-dimensional
surfaces of Amatatsu et al. \cite{am96}, where the spin-orbit 
configuration interaction method was used with a better 
basis by changing the valence double-$\zeta$ level to the
triple-$\zeta$ one. Out of the nine coordinates of the surfaces,   
the six coordinates which are neglected in the simulations were   
fixed at their equilibrium values \cite{ru09}.  

The electric-dipole moment functions ($d(R,r,\theta)$) coupling radiatively the  
$\tilde{X}^1A_1$ ground state with $^3Q_0$ and $^1Q_1$
were obtained from the {\it ab initio} 
calculations of Alekseyev et al. \cite{alpri,al07II}, subsequently fitted to analytical forms 
used in the simulations \cite{ru09}.

\subsection{Time-dependent quantum calculations}
\label{V.B}

CH$_3$I is initially in the rovibrational ground state $\phi_0(R,r,\theta)$, variationally
obtained from Eq.~\eqref{3} within the framework of an adiabatic approximation detailed
in ref.~\cite{Alberto1}. CH$_3$I is then 
excited to one of the $^3Q_0$ and $^1Q_1$ electronic 
states, creating a wave packet $\Phi_0(R,r,\theta)$ (see Eq.~\eqref{5a}) undergoing
dynamical evolution according to Eq.~\eqref{DD.7}.
In order to solve this equation, the wave packet $\Phi_t(R,r,\theta)$ 
is represented in a basis set consisting of a two-dimensional rectangular
grid for the radial coordinates and an angular basis including $24$
Legendre polynomials for the $\theta$ coordinate. The rectangular grid
consists of $450$ equidistant points in the $R$ coordinate in the
range $3.5 \, a_0 \leq R \leq 16.0 \, a_0$, and $32$ equidistant points
in the $r$ coordinate in the range $-1.6 \, a_0 \leq r \leq 1.6 \, a_0$.
Propagation of the wave packet is performed by representing the evolution
operator by means of a Chebychev polynomial expansion. The wave packet
was propagated for $200$ fs with a time step $\Delta t = 0.4$ fs, and   
was absorbed at the edge of the grid in the $R$ coordinate after  
each propagation time step by multiplying it by   
the function $exp[- \alpha (R-R_{abs})^2]$, with $\alpha=0.9$ 
$a_0^{-2}$ and $R_{abs}=13.0$ $a_0$. In order to obtain the 
product fragment distributions of interest, the wave packet was projected 
out in the asymptotic region onto the fragment states. Details of the 
projection procedure are given in ref.~\cite{Alberto1}.

\subsection{Comparison between semiclassical and quantum results}
\label{V.C}

$\Phi_0(R,r,\theta)$ was expressed as a product of three independent Gaussians respectively depending on $R$,
$r$ and $\theta$, a very good approximation in the present case. As in section~\ref{II.D}, the density 
$\rho_0(\mathbf{\Gamma_0})$ was then deduced from Eq.~(5.20) of ref.~\cite{Schinke}.  

The results obtained by means of the forward I, forward II and Goursaud \emph{et al.} methods involved
5 million, 100 thousand and 1 million trajectories, respectively. The results obtained by means of the backward
method involved 60 thousand trajectories per vibrational level (see end of section~\ref{III}). These numbers are
for each electronic state, $^3Q_0$ and $^1Q_1$. For the backward approach, $\eta_r$ and $\eta_\theta$ were kept
at 0.02 while $\eta_p$ and $\eta_{P_{\theta}}$ were taken at 0.2 (in atomic units for $\eta_r$, $\eta_p$ and 
$\eta_{P_{\theta}}$ and in radian for $\eta_\theta$). $R_f$ was found to be equal to 13 and 10 bohr for the 
$^3Q_0$ and $^1Q_1$ states, respectively.

\begin{figure}
 \includegraphics[width=180mm]{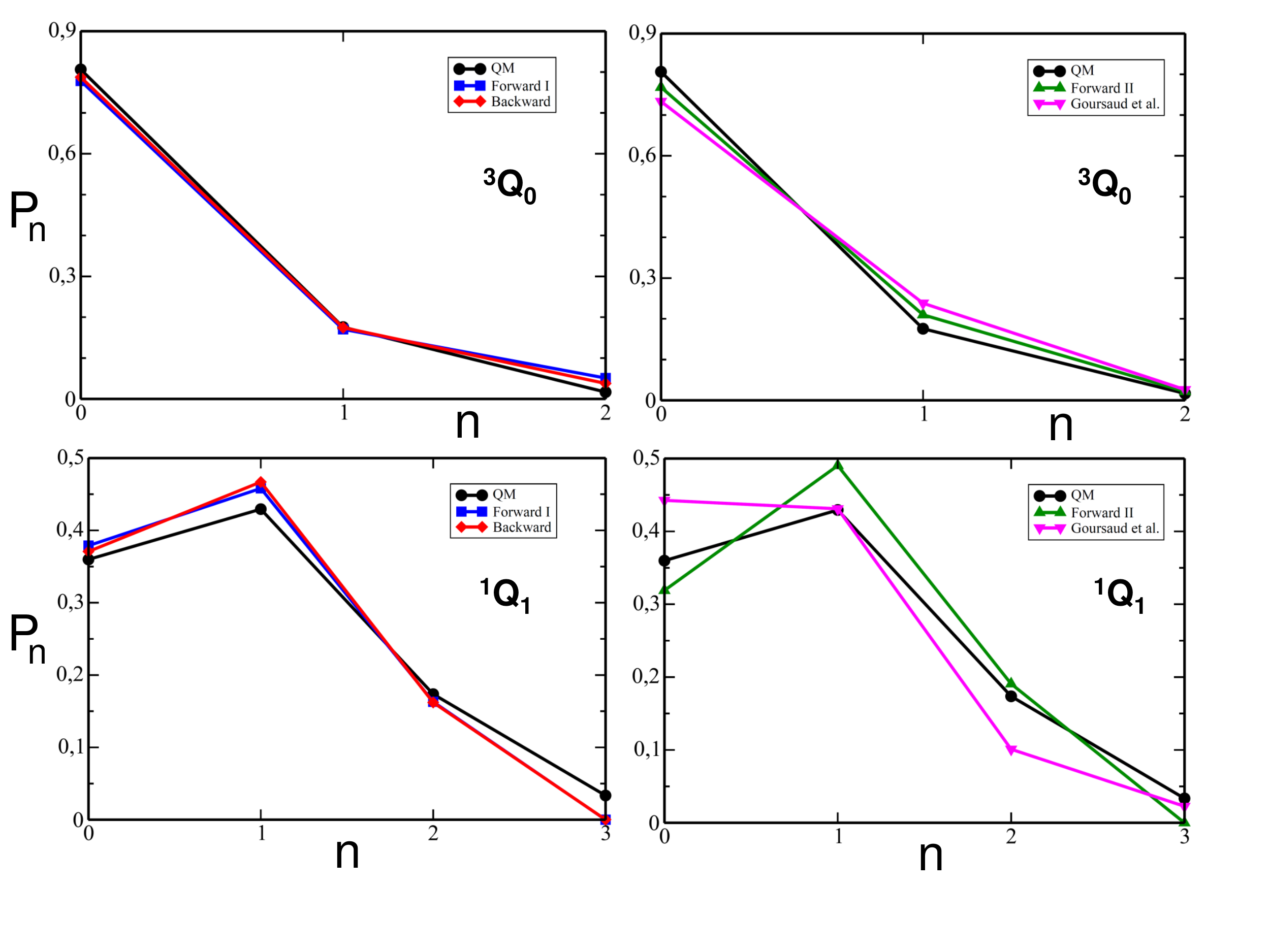}
 \caption{Vibrational state distributions in the $^3Q_0$ and $^1Q_1$ states 
 found by means of the QM, forward I, forward II, backward and Goursaud \emph{et al.}
  methods.   
 \label{figure6}}
\end{figure}

\begin{figure}
 \includegraphics[width=180mm]{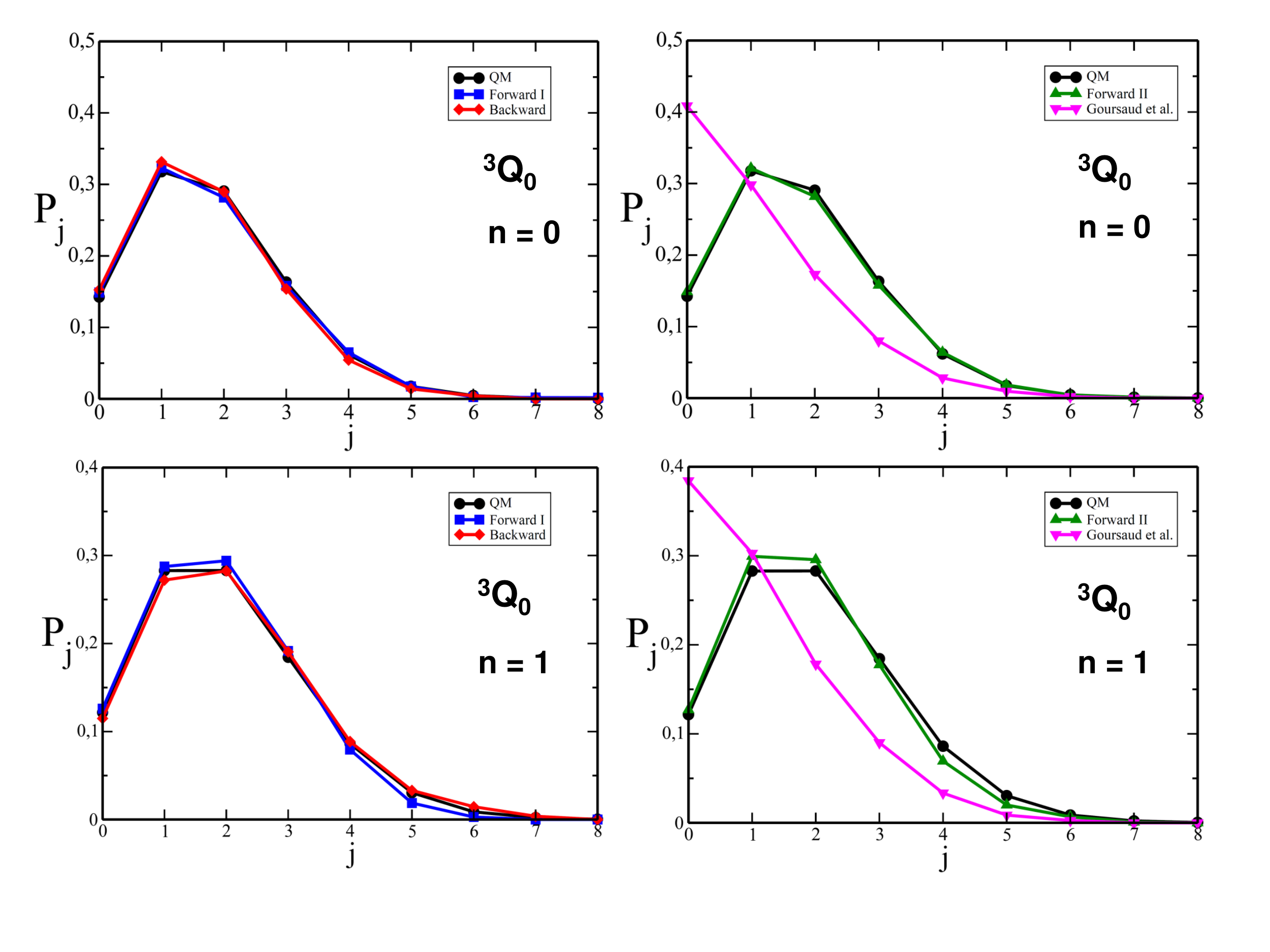}
 \caption{Vibrationally resolved rotational state distributions in the $^3Q_0$ state
 found by means of the QM, forward I, forward II, backward and Goursaud \emph{et al.}
  methods.   
 \label{figure7}}
\end{figure}

\begin{figure}
 \includegraphics[width=180mm]{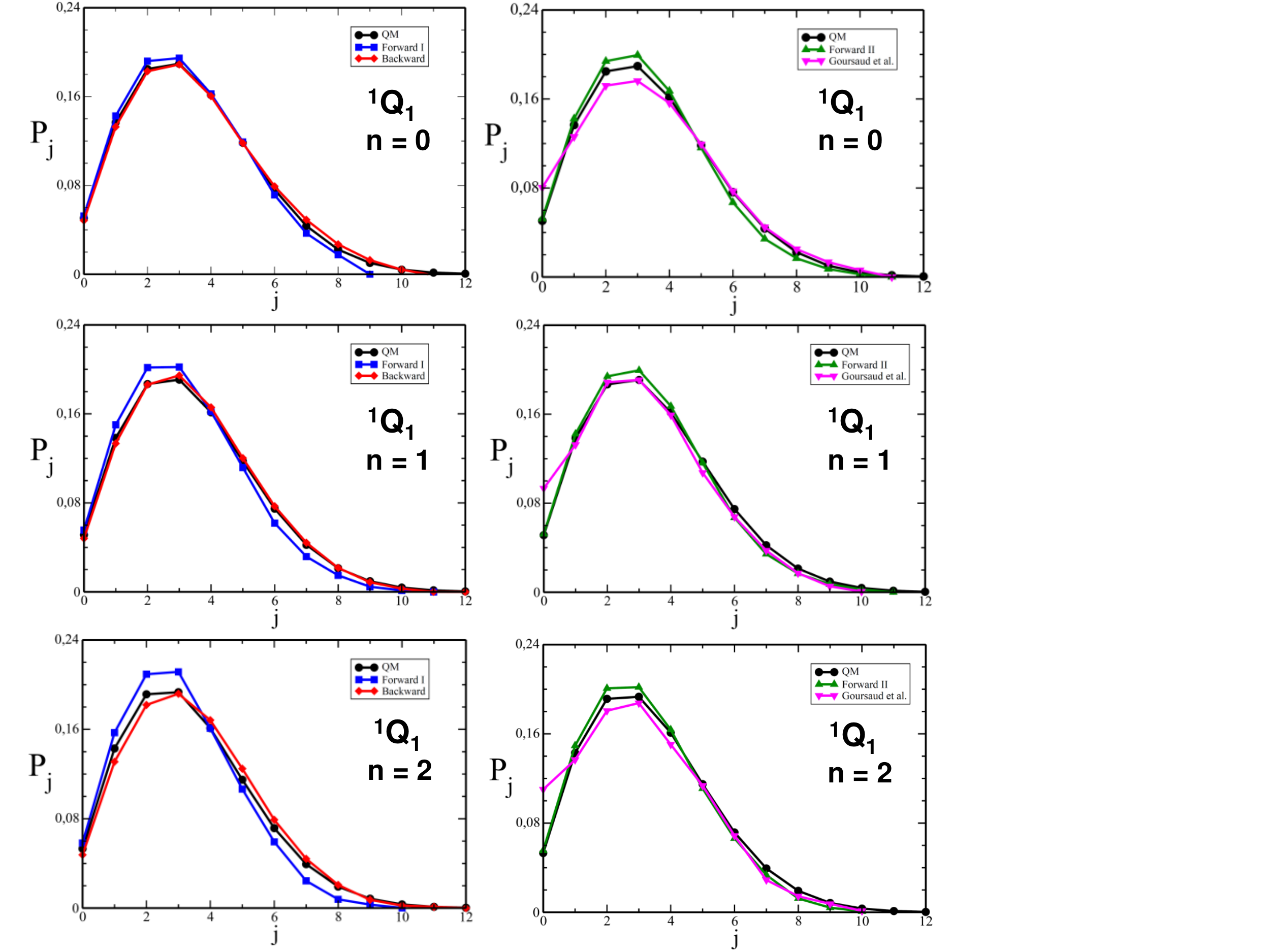}
 \caption{Vibrationally resolved rotational state distributions in the $^1Q_1$ state
 found by means of the QM, forward I, forward II, backward and Goursaud \emph{et al.}
  methods.    
 \label{figure8}}
\end{figure}

The vibrational state populations $P_n$, deduced from Eq.~\eqref{1a} by summing over $j$, are given in Fig.~\ref{figure6}. 
The vibrationally resolved rotational state distributions are displayed in Fig.~\ref{figure7} and Fig.~\ref{figure8}. 
The latter are simply denoted $P_j$, but the value of $n$ to which they refer is indicated.   

The agreement between backward and quantum mechanical (QM) results is quantitative for all the distributions but
$P_n$ in the $^1Q_1$ state, for which it is nevertheless very good. 

These conclusions hold between forward I and QM results, apart from $P_j$, $n=1,2$, in the $^1Q_1$ state for which 
the quality of the agreement decreases. The reason seems to be the following. As seen at the end of section~\ref{II.D},
the partial cross section involves a delta function through $\rho_{tr}(R_f,P_f)$ (see Eqs.~\eqref{31} and~\eqref{13})
which is replaced by a narrow bin. However, the bin cannot be too narrow, for a negligible amount of trajectories
would contribute to the partial cross sections. With a true delta function, $P$ would be strictly equal to $\hbar k_{nj}$.
With a bin, however, there is an uncertainty on the value of $P$ around $\hbar k_{nj}$, which seems to cause the 
differences observed between the forward I results and the backward and quantum ones. By dividing the width of the bin by two,
we indeed reduced the differences. We tried to reduce even more the width of the bin, but this prevented from converging
the partial cross sections. 
The minimum number of trajectories necessary to converge the calculations was found to be 800 thousand for the forward I
method in both electronic states, against 10 thousand for $n=0$ in the $^3Q_0$ state, 20 thousand for $n=1$ in the $^3Q_0$ state,
8 thousand for $n=0$ in the $^1Q_1$ state, 10 thousand for $n=1$ in the $^1Q_1$ state and 30 thousand for $n=2$ in the $^1Q_1$ state.
As a matter of fact, the backward method is at the same time more efficient and more accurate than the forward I method.

The agreement between forward II and QM results is surprisingly good for all the rotational distributions
(we have no explanation for that), a bit less
for the vibrational ones, especially in the $^1Q_1$ state. Like the forward I and backward methods, the forward II 
method accounts for the vibrational inversion in the $^1Q_1$ state. 

The agreement between Goursaud \emph{et al.} and QM results is very satisfying for the rotational distributions
in the $^1Q_1$ state, except for $j=0$, and for the vibrational distribution in the $^3Q_0$ state. 
On the other hand, the method of Goursaud \emph{et al.} fails at reproducing the vibrational inversion in the $^1Q_1$ state,  
and badly describes the rotational distributions in the $^3Q_0$ state, even qualitatively. Overall, this method
is only semi-quantitative. 

As previously stated, the energy available to the final products is larger in
the $^1Q_1$ than in the $^3Q_0$ state. This explains in part why there is more vibrational and rotational
excitation in $^1Q_1$ than in $^3Q_0$. The system is thus less quantum-like in $^1Q_1$ than in $^3Q_0$,
justifying thereby why Goursaud \emph{et al.} method, the more classical of the four semiclassical approaches
considered in this work, is able to reproduce the energy partitioning in $^1Q_1$ and not in $^3Q_0$.

\section{Conclusion}
\label{VI}

In their concluding remarks, Brown and Heller \cite{Heller2} raised the basic issue of including rotations
in their semiclassical Wigner description of photodissociation dynamics so as to make it full-dimensional and thus, realistic. 
This issue has been solved in the present work. 

Three methods have been proposed, respectively called forward I, forward II and backward. While forward approaches
involve trajectories run from the reagent molecule onto the products, the backward one deals with trajectories
run in the reverse direction. This makes the backward method state-selective, and thus, ideally suited to the description 
of state-correlated distributions measured in velocity imaging experiments.

The forward I and backward methods are exact applications of the semiclassical Wigner treatment.
They closely parallel the quantum description as far as possible.
The forward II approach involves an approximation strongly increasing the numerical efficiency of the 
semiclassical Wigner treatment as compared to the forward I method. 

In addition to the usual vibrational Wigner distributions, these three approaches include rotational Wigner distributions
which seem to be introduced for the first time in the present context. These densities have remarkable structures clearly showing that classical trajectories most contributing to rotational state $j$ are those reaching the products with a rotational angular momentum close to $[j(j+1)]^{1/2}$ (in $\hbar$ unit).  

The results obtained by means of these methods are compared with rigorous quantum results in the case of Guo's  
triatomic-like model of methyl iodide photodissociation \cite{Guo2} and very good - often quantitative - agreement is found, 
especially with
the forward I and backward methods. In comparison, the standard and more classical method of Goursaud \emph{et al.} 
\cite{Goursaud} is only semi-quantitative. Last but not least, the backward approach appears to be 
much more powerful, and even more accurate than the forward I method which requires far more
trajectories than the former approach to provide converged results.

This study demonstrates the applicability of the semiclassical Wigner treatment to realistic triatomic photodissociations 
and confirms its level of accuracy as compared to the initial work of Brown and Heller \cite{Heller2}. 
%way is open towards several 
Important next steps are the extention of the method to processes involving non adiabatic transitions and/or polyatomic
species.

\newpage

\renewcommand{\theequation}{A.\arabic{equation}}

\setcounter{equation}{0}

\section*{Appendix A: Derivation of Eq.~\eqref{2}}

Consider a given state $\Psi^{JM\epsilon}(\mathbf{R},\mathbf{r})$ of ABC for the value $J$ of the total angular momentum
quantum number, the value $M$ of its projection on the $Z$-axis of the laboratory reference frame, and the parity $\epsilon$
under inversion of $\mathbf{R}$ and $\mathbf{r}$. This state can generally be expanded as \cite{Octavio2,Balint3} 
\\
\begin{equation}
\bar{\Psi}^{JM\epsilon}(\mathbf{R},\mathbf{r}) = \sum_{\Omega}\;W^{J\epsilon}_{M\Omega}(\alpha,\beta,\gamma)
\;\frac{\Psi^{J\epsilon}_{M\Omega}(R,r,\theta)}{Rr}
\label{AA.1}
\end{equation}
with
\begin{equation}
W^{J\epsilon}_{M\Omega}(\alpha,\beta,\gamma)=
\Big[\frac{2J+1}{16\pi^2(1+\delta_{\Omega0})}\Big]^{1/2}\;[D^{J}_{M,\Omega}(\alpha,\beta,\gamma)+\epsilon (-1)^{(J+\Omega)}
D^{J}_{M,-\Omega}(\alpha,\beta,\gamma)].
\label{AA.2}
\end{equation}
\\
$\alpha$, $\beta$ and $\gamma$ are the Euler angles orienting ABC in the laboratory reference frame \cite{Edmonds}. 
The helicity quantum number $\Omega$ is the projection of the total angular momentum on $\mathbf{R}$, chosen as the $Z$-axis of the 
body-fixed frame. $D^{J}_{M,\Omega}(\alpha,\beta,\gamma)$ and $D^{J}_{M,-\Omega}(\alpha,\beta,\gamma)$ are Wigner D-matrix elements
\cite{Edmonds}.

When $J=0$, $M=\Omega=0$ and $\epsilon$ is necessarily equal to 1 for $W^{J\epsilon}_{M\Omega}(\alpha,\beta,\gamma)$ no to be 0.
Given that $D^{0}_{0,0}(\alpha,\beta,\gamma)=1$ (see Eq. (4.1.26) of \cite{Edmonds}), Eq.~\eqref{AA.1} reduces to
\\
\begin{equation}
\bar{\Psi}^{001}(\mathbf{R},\mathbf{r}) = \frac{\Psi^{01}_{00}(R,r,\theta)}{8^{1/2}\pi R r}.
\label{AA.3}
\end{equation}
\\
We can thus rewrite $\bar{\Psi}_E^{nj}(\mathbf{R},\mathbf{r})$ and $\bar{\Phi}_t(\mathbf{R},\mathbf{r})$ in Eq.~\eqref{DD.4} as
\\
\begin{equation}
\bar{\Psi}_E^{nj}(\mathbf{R},\mathbf{r}) = \frac{{\Psi_E^{nj}}(R,r,\theta)}{8^{1/2}\pi Rr}
\label{AA.3b}
\end{equation}
and
\begin{equation}
\bar{\Phi}_t(\mathbf{R},\mathbf{r}) = \frac{\Phi_t(R,r,\theta)}{8^{1/2}\pi Rr}
\label{AA.3a}
\end{equation}
\\
where for convenience's sake, subscripts and indices relative to angular momentum quantum numbers and parity have been dropped.

Eqs.~\eqref{DD.4},~\eqref{AA.3b} and \eqref{AA.3a} lead to 
\\
\begin{equation}
\Sigma_E^{nj} = \left| \int\;\frac{\mathbf{dR} \mathbf{dr}}{8\pi^2 R^2 r^2}\;{\Psi_E^{nj}}{}^*(R,r,\theta)\;
\Phi_t(R,r,\theta) \right|^2.
\label{AA.4}
\end{equation}
\\
Moreover, some steps of algebra allow to prove the two identities 
\\
\begin{equation}
\mathbf{dR} \mathbf{dr}=R^2r^2dRdr sin\theta d\theta d\alpha sin\beta d\beta d\gamma
\label{AA.5}
\end{equation}
and
\begin{equation}
\int\; d\alpha sin\beta d\beta d\gamma = 8\pi^{2}.
\label{AA.6}
\end{equation}
\\
Eqs.~\eqref{AA.4}-\eqref{AA.6} finally lead to Eq.~\eqref{2}. 

For $J=0$, $\bar{\Psi}_E^{nj}(\mathbf{R},\mathbf{r})$ and $\bar{\Phi}_t(\mathbf{R},\mathbf{r})$ are shown to satisfy the 
Schr\"{o}dinger equations \cite{Schatz,Tennyson}
%\begin{equation}
%\hat{\bar{H}}_g=[\hat{\bar{T}}+\hat{V}_g] \bar{\phi}_0 = E_0 \bar{\phi}_0 
%\label{AA.7}
%\end{equation}
\\
\begin{equation}
\hat{\bar{H}}_e=[\hat{\bar{T}}+\hat{V}_e] \bar{\Psi}_E^{nj} = E \bar{\Psi}_E^{nj} 
\label{AA.8}
\end{equation}
and
\begin{equation}
i\hbar \frac{d\bar{\Phi}_t}{dt}=\hat{\bar{H}}_e \bar{\Phi}_t,
\label{DD.5}
\end{equation}
\\
with
\begin{equation}
\hat{\bar{T}} = - \frac{\hbar^2}{2\mu R^2}\frac{\partial}{\partial R}R^2\frac{\partial}{\partial R}
- \frac{\hbar^2}{2m r^2}\frac{\partial}{\partial r}r^2\frac{\partial}{\partial r}
- \frac{\hbar^2}{2Isin\theta}\frac{\partial}{\partial \theta}sin\theta\frac{\partial}{\partial \theta}.
\label{AA.9}
\end{equation}
\\
The reduced moment of inertia $I$ satisfies the identity
\\
\begin{equation}
\frac{1}{I} = \frac{1}{\mu R^2}+\frac{1}{m r^2}.
\label{AA.10}
\end{equation}
\\
From Eqs.~\eqref{AA.3b},~\eqref{AA.3a},~\eqref{AA.8}-\eqref{AA.9},
${\Psi}_E^{nj}(R,r,\theta)$ and ${\Phi}_t(R,r,\theta)$ are shown to satisfy Eqs.~\eqref{4}-\eqref{5}.

%One might be tempted by the following straightforward but wrong derivation of Eq.~\eqref{9}. 
%Using Eq.~\eqref{8}, Eq.~\eqref{2} reads
%\\
%\begin{equation}
%\Sigma_E^{nj} = \Big| \int\;dR dr d\theta\;sin\theta\;{\Psi_E^{nj}}^*(R,r,\theta)\;\Phi_0(R,r,\theta) \Big|^2.
%\label{DD.8}
%\end{equation}
%\\ 
%Now, writing the identity operator as
%\begin{equation}
%\hat{1} = e^{i\hat{H}_e t/\hbar}e^{-i\hat{H}_e t/\hbar}
%\label{DD.9}
%\end{equation}
%\\
%and inserting it between the two states in Eq.~\eqref{DD.8} might seem to justify Eq.~\eqref{9}. However, this is 
%untrue because propagator $e^{i\hat{H}_e t/\hbar}$ acts on its left on $sin\theta\;{\Psi_E^{nj}}^*(R,r,\theta)$ instead of %${\Psi_E^{nj}}^*(R,r,\theta)$ alone. In fact, the Achille's heel of this demonstration is that Eq.~\eqref{DD.9} does not seem to be %correct. Such a kind of problem does not arise within the framework of Cartesian coordinates, 
%as seen in the previous demonstration of Eq.~\eqref{9}.

\newpage

\renewcommand{\theequation}{B.\arabic{equation}}

\setcounter{equation}{0}

\section*{Appendix B: Derivation of Eq.~\eqref{12}}

Any overlap $\Sigma$ defined by 
\begin{equation}
\Sigma = \left|\int\;dR dr d\theta \; \Psi_1^*(R,r,\theta) \; \Psi_2(R,r,\theta)\right|^2
\label{BB.1}
\end{equation}
can be rewritten as \cite{Heller1,Heller2,Schinke2}
\begin{equation}
\Sigma = (2\pi \hbar)^3\int\;dR dr d\theta dP dp dP_{\theta} \; \rho_1(R,r,\theta,P,p,P_{\theta}) \; \rho_2(R,r,\theta,P,p,P_{\theta})
\label{BB.2}
\end{equation}
\\
where $\rho_l(R,r,\theta,P,p,P_{\theta})$, $l=1$ or 2, is the Wigner density defined by
\\
\begin{equation}
\rho_l(R,r,\theta,P,p,P_{\theta}) = \frac{1}{(\pi \hbar)^3}\int\;ds_R ds_r ds_{\theta} \; 
e^{2i(Ps_R+ps_r+P_{\theta}s_{\theta})/\hbar}\;
\Psi_l^*(R+s_R,r+s_r,\theta+s_{\theta})\;
\Psi_l(R-s_R,r-s_r,\theta-s_{\theta}).
\label{BB.3}
\end{equation}
\\
This expression is a generalization of Eq.~\eqref{I1} to three dimensions.
 
A pedestrian demonstration of the strict equivalence between Eqs.~\eqref{BB.1} and \eqref{BB.2} 
for one configuration space coordinate is given in Appendix B of ref.~\cite{Part3}. 
In the present case of three coordinates, the developments are more tedious, but present no difficulty. 

Setting
\begin{equation}
\Psi_1(R,r,\theta) = \Big[\frac{\mu}{2\pi\hbar^2 k_{nj}}\Big]^{1/2}e^{ik_{nj}R}\;\chi_n(r)\;sin\theta\; Y_{j}^0(\theta)
\label{BB.4}
\end{equation}
and
\begin{equation}
\Psi_2(R,r,\theta) = \Phi_t(R,r,\theta),
\label{BB.5}
\end{equation}
\\
we arrive from Eqs.~\eqref{BB.2} and \eqref{BB.3} at  
\\
\begin{equation}
\Sigma_E^{nj} = (2\pi \hbar)^3
\lim\limits_{t \to +\infty} 
\int\;dR dr d\theta dP dp dP_{\theta}\; \rho_t(R,r,\theta,P,p,P_{\theta})
\; \rho_{tr}(R,P)\; \rho_n(r,p)\; \rho_j(\theta,P_{\theta}).
\label{BB.6}
\end{equation}
\\
$\rho_t(R,r,\theta,P,p,P_{\theta})$ is related to $\Phi_t(R,r,\theta)$ by Eq.~\eqref{BB.3}. 
The translational Wigner distribution $\rho_{tr}(R,P)$ is given by
\\
\begin{equation}
\rho_{tr}(R,P) = \frac{1}{\pi \hbar}\int\;ds \; 
e^{2iPs/\hbar}\;
\Big[\frac{\mu}{2\pi\hbar^2 k_{nj}}\Big]\;
e^{-ik_{nj}(R+s)}\;
e^{ik_{nj}(R-s)}
\label{BB.7}
\end{equation}
\\
while the vibrational and rotational Wigner distributions $\rho_n(r,p)$ and $\rho_j(\theta,P_{\theta})$
are given by Eqs.~\eqref{14} and \eqref{15}. Eq.~\eqref{BB.7} gives
\\
\begin{equation}
\rho_{tr}(R,P) = \Big[\frac{\mu}{2\pi\hbar^2 k_{nj}}\Big]\;
\frac{1}{\pi \hbar}\int\;ds \; e^{2i(P-\hbar k_{nj})s/\hbar}
\label{BB.8}
\end{equation}
which, using 
\begin{equation}
\delta(x) = \frac{1}{2\pi}\;
\int\;ds \; e^{isx},
\label{BB.9}
\end{equation}
\\
leads to
\begin{equation}
\rho_{tr}(R,P) = \Big[\frac{\mu}{2\pi\hbar^2 k_{nj}}\Big]\;
\frac{2}{\hbar}\ \delta[2(P-\hbar k_{nj})/\hbar]
\label{BB.10}
\end{equation}
\\
or equivalently, Eq.~\eqref{13}. Eq.~\eqref{BB.10} is indeed readily obtained from Eq.~\eqref{13} by means of the following theorem
\\
\begin{equation}
\delta[f(x)] = \sum_k\;\frac{1}{|f'(x_k)|} \delta(x-x_k)
\label{BB.11}
\end{equation}
\\
where the $x_k$'s are solutions of $f(x)=0$ \cite{CDL}.

\newpage

\renewcommand{\theequation}{C.\arabic{equation}}

\setcounter{equation}{0}

\section*{Appendix C: Semiclassical limit of the rotational Wigner distribution}

\begin{figure}
 \includegraphics[width=200mm]{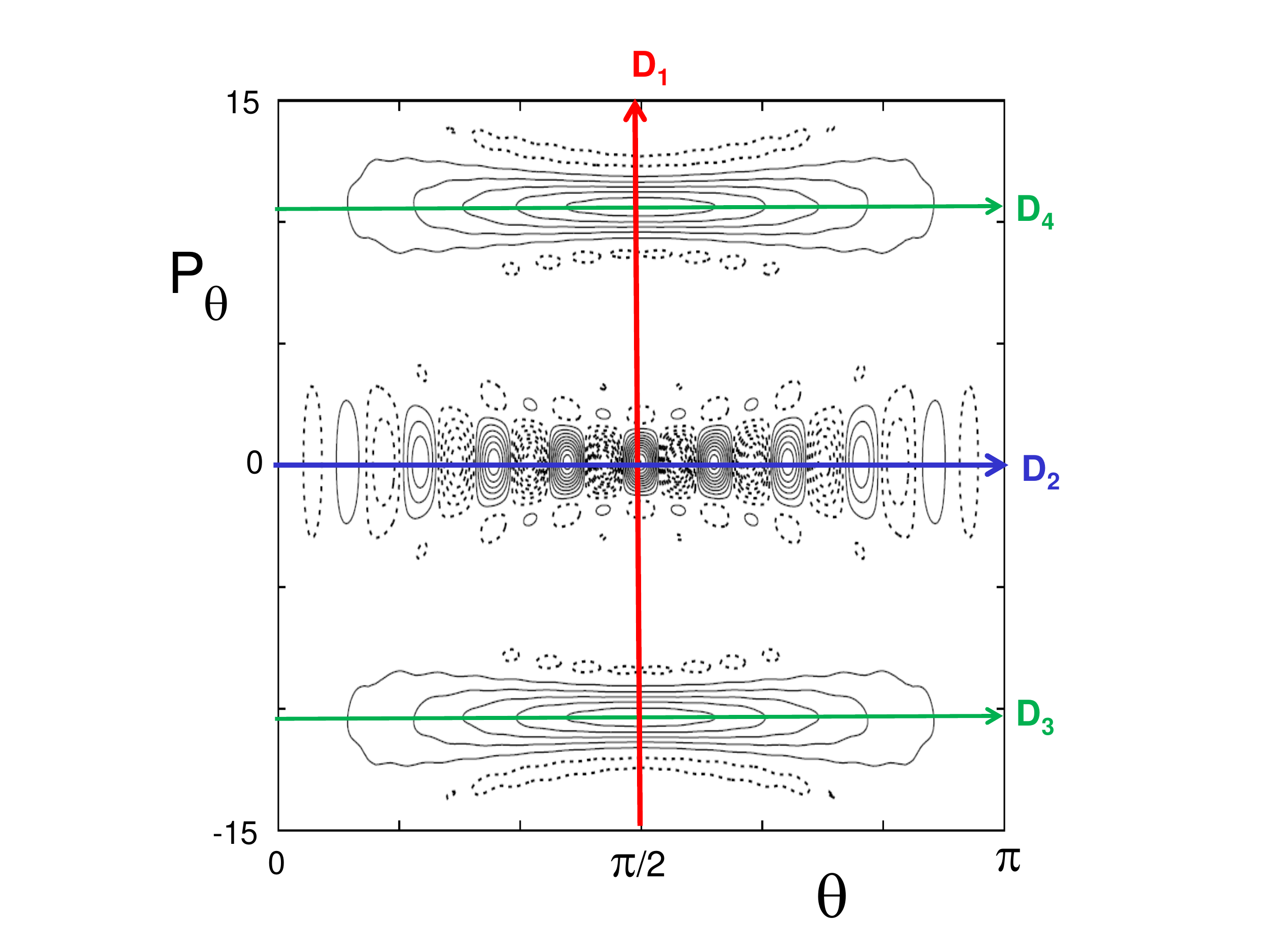}
 \caption{Contour plot representation of $\rho_{10}(\theta,P_{\theta})$ and directions $D_i$, $i=1-4$, along which semiclassical expressions of the rotational Wigner distribution are derived in Appendix C. $P_{\theta}$ is in $\hbar$ unit.
 Solid and dashed contours correspond to positive and negative densities, respectively.  
 \label{figureC1}}
\end{figure}

Our goal here is to derive an analytical expression of $\rho_j(\theta,P_{\theta})$ within the semiclassical 
approximation along the relevant directions defined by $\theta=\pi/2$ and $P_{\theta} = 0,\pm(j+1/2)$. 
$P_{\theta}$ is expressed in $\hbar$ unit in throughout this appendix. These directions
are emphasized in Fig.~\ref{figureC1} for $j=10$. The first two ones (D$_1$ and D$_2$)
are the two orthogonal symmetry axis of $\rho_j(\theta,P_{\theta})$. The next two ones (D$_3$ and D$_4$) 
are not exact symmetry axis, but
they can be considered as local symmetry axis of the lateral ridges for sufficiently large $j$'s, as shown
further below.

\begin{figure}
 \includegraphics[width=200mm]{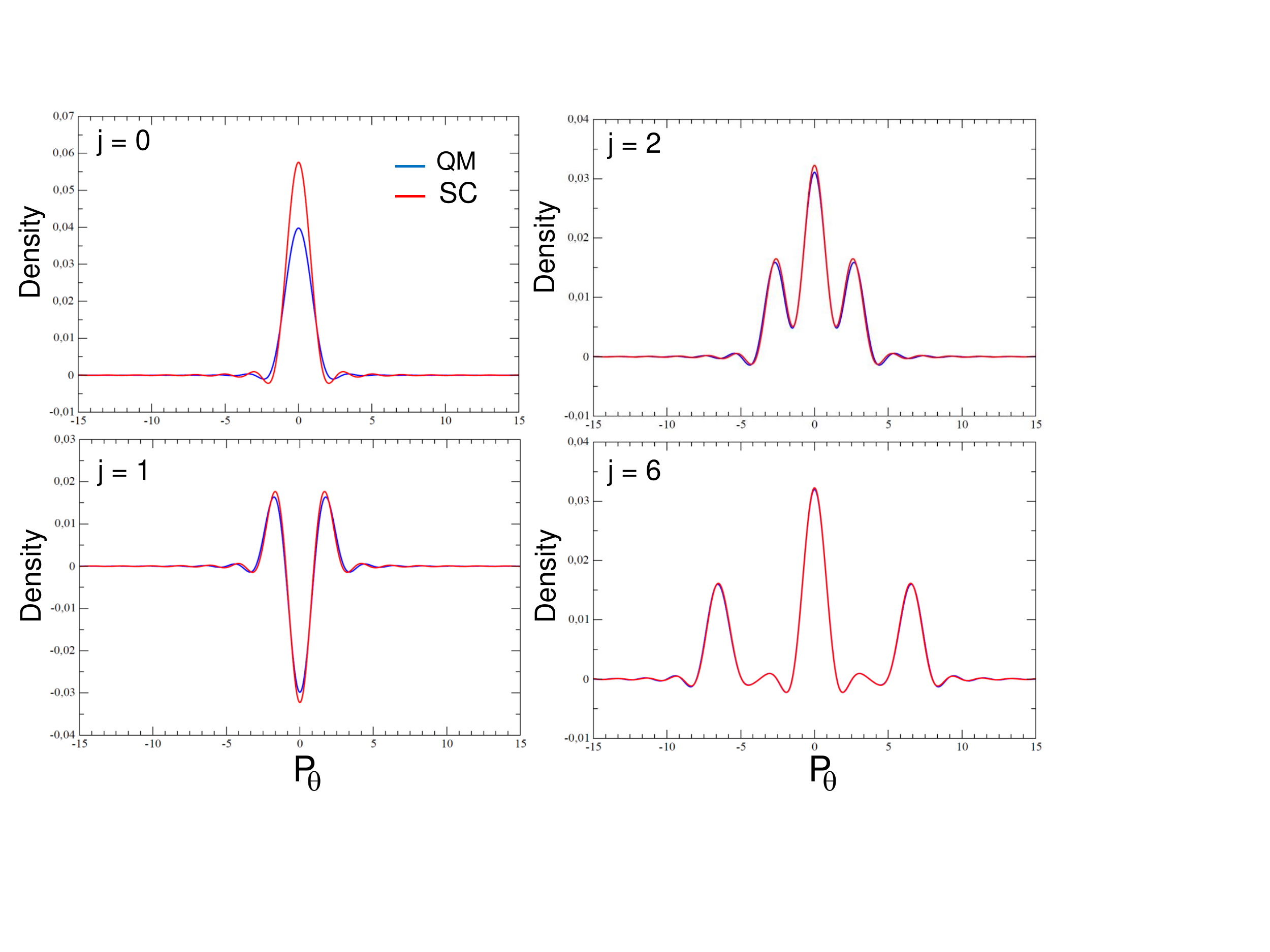}
 \caption{Comparison between the semiclassical value of $\rho_j(\pi/2,P_{\theta})$ along the $D_1$ direction
 (see Fig.~\ref{figureC1}), obtained from Eq.~\eqref{CC.7},
and its exact quantum value for $j=0,1,2$ and 6. $P_{\theta}$ is in $\hbar$ unit.
 \label{figureC2}}
\end{figure}

The semiclassical (WKB) limit of the Legendre polynomial $P_j(cos\theta)$ is given by \cite{More}
\\
\begin{equation}
P_j(cos\theta) = \frac{2 cos[(j+1/2)\theta-\pi/4]}{[2\pi(j+1/2)sin\theta]^{1/2}}.
\label{CC.1}
\end{equation}
\\
From Eqs.~\eqref{7},~\eqref{15} and~\eqref{CC.1}, we have
\\
\begin{equation}
\rho_j(\theta,P_{\theta}) = \frac{1}{\pi^3}\int\;ds \; 
e^{2iP_{\theta}s}\;
[sin(\theta+s) sin(\theta-s)]^{1/2}\;cos[(j+1/2)(\theta+s)-\pi/4]\;cos[(j+1/2)(\theta-s)-\pi/4].
\label{CC.2}
\end{equation}
\\
Using the fact that 
\\
\begin{equation}
cosx=\frac{e^{ix}+e^{-ix}}{2},
\label{CC.3}
\end{equation}
we arrive after some steps of simple algebra at
\\
\begin{equation}
\rho_j(\theta,P_{\theta}) = \frac{1}{2\pi^3}\int\;ds \; 
e^{2iP_{\theta}s}\;
[sin(\theta+s) sin(\theta-s)]^{1/2}\;\big[cos[(2j+1)s]+sin[(2j+1)\theta]\big].
\label{CC.4}
\end{equation}
\\
Along direction D$_1$, defined by $\theta=\pi/2$, we have
\\
\begin{equation}
\rho_j(\pi/2,P_{\theta}) = \frac{1}{2\pi^3}\int^{\pi/2}_{-\pi/2}\;ds \; 
e^{2iP_{\theta}s}\;
coss\;\big[cos[(2j+1)s]+(-1)^j\big].
\label{CC.5}
\end{equation}
\\
The boundaries of the integral are determined by Eq.~\eqref{15a}. From Eq.~\eqref{CC.3} and given that
\\
\begin{equation}
\int^{\pi/2}_{-\pi/2} ds\;e^{2ias} = \frac{sin(\pi a)}{a}=\pi sinc(\pi a), 
\label{CC.6}
\end{equation}
we finally obtain
%\\
%\begin{equation}
%\begin{split}
%h_j^{i+1}
%& = a \\
%& \quad - \left. b\right. \\
%& \quad -\left. c\right.
%\end{split}
%\label{CC.9}
%end{equation}
\\
\begin{equation}
\begin{split}
\rho_j(\pi/2,P_{\theta})
& = \frac{1}{4\pi^2}(-1)^j\;\big[sinc[\pi (P_{\theta}+1/2)]+sinc[\pi (P_{\theta}-1/2)]\big] \\
& \quad + \left. \frac{1}{8\pi^2}\big[sinc[\pi (P_{\theta}+j+1)]+sinc[\pi (P_{\theta}+j)]\right. \\
& \quad + \left. sinc[\pi (P_{\theta}-j)]+sinc[\pi (P_{\theta}-j-1)]\big]\right..
\end{split}
\label{CC.7}
\end{equation}
\\
This expression is represented in Fig.~\ref{figureC2} for $j=0,1,2$ and 6 and is compared with the exact Wigner distribution.
For $j=0$, the agreement is already satisfying. For $j=1$ and 2, it is very good and for $j=6$, there is virtually no
difference between both results. 

For large $j$'s, the central peak is due to the first pair of sinc functions in Eq.~\eqref{CC.7}
and its sign is thus given by $(-1)^j$.
The lateral peaks, or cuts of ridges, centered at $-(j+1/2)$ and $(j+1/2)$ are due to the second and third pairs of sinc functions, respectively.
They are always positive. Each of these three peaks is thus made of two narrower peaks sufficiently close to each other for 
resulting in a single peak. The full width at half maximum is 1.2 for the sinc function and 1.64 for the central and lateral 
peaks.  

For small $j$'s, the situation is a bit more complex, since the three peaks overlap. Consequently, the lateral peaks
are not exactly centered at $-(j+1/2)$ and $(j+1/2)$.  
For $j=0$, the lateral peaks have merged with the central peak, and their top is rigorously at 0.   
 
Along direction D$_2$, defined by $P_{\theta}=0$, we have from Eq.~\eqref{CC.4}
\\
\begin{equation}
\rho_j(\theta,0) = \frac{1}{2\pi^3}\int_{s_-}^{s_+}\;ds \; 
[sin(\theta+s) sin(\theta-s)]^{1/2}\;\big[cos[(2j+1)s]+sin[(2j+1)\theta]\big]
\label{CC.8}
\end{equation}
\\
where the boundaries are found from Eq.~\eqref{15a} to be given by $s_{\pm}=\pm\theta$ if $\theta \le \pi/2$
and $s_{\pm}=\pm (\pi-\theta)$ if $\theta > \pi/2$. It is thus clear that the angular range around $\pi/2$ mostly
contributes to $\rho_j(\theta,0)$.

\begin{figure}
 \includegraphics[width=200mm]{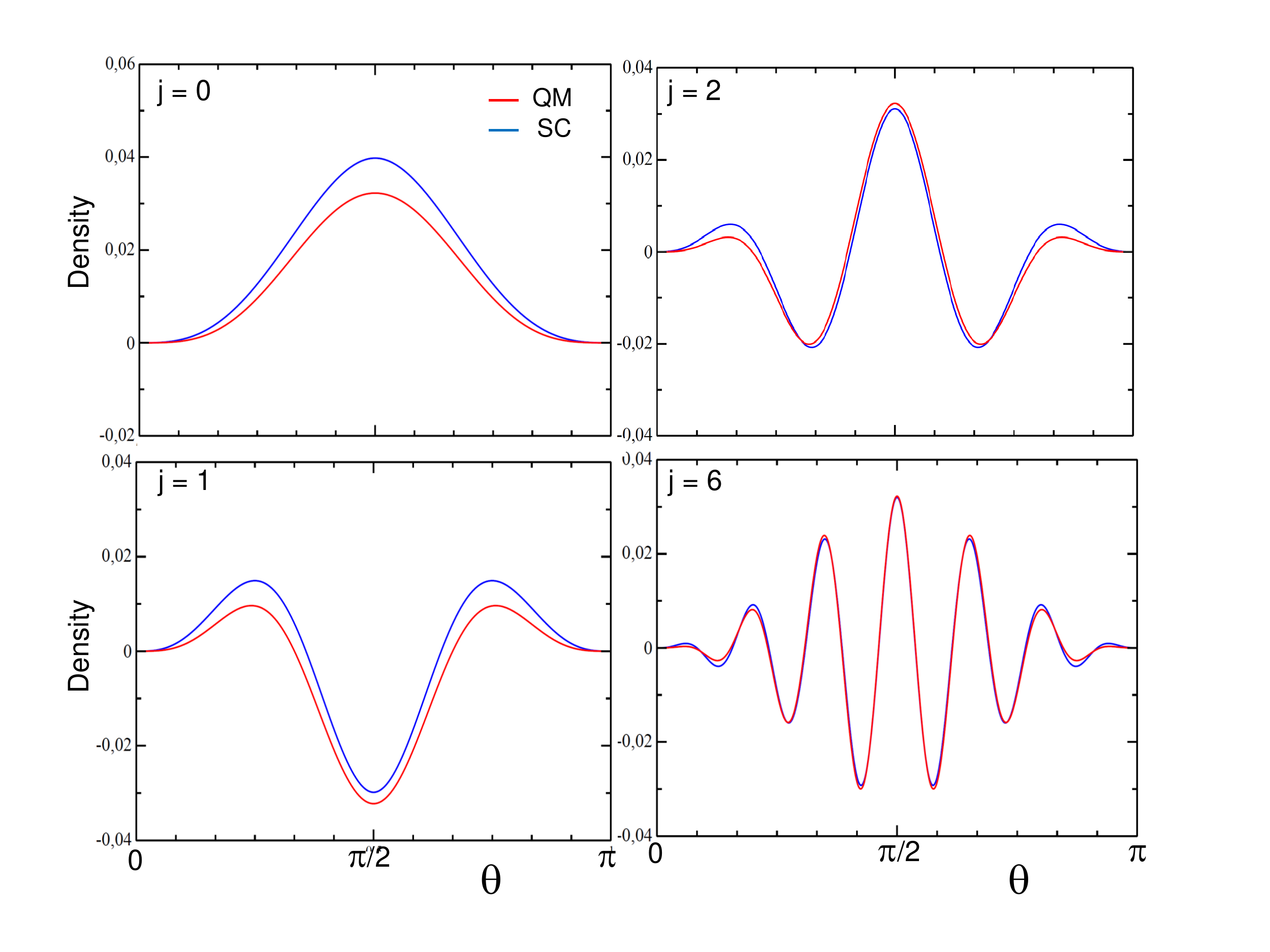}
 \caption{Comparison between the semiclassical value of $\rho_j(\theta,0)$ along the $D_2$ direction
 (see Fig.~\ref{figureC1}), obtained from Eq.~\eqref{CC.10},
and its exact quantum value for $j=0,1,2$ and 6.
 \label{figureC3}}
\end{figure}

For significantly large $j$'s, $cos[(2j+1)s]$ strongly varies within the previous range and can thus be neglected 
with respect to $sin[(2j+1)\theta]$. We thus arrive at 
\\
\begin{equation}
\rho_j(\theta,0) = \frac{1}{2\pi^3}sin[(2j+1)\theta]\big]\int_{s_-}^{s_+}\;ds \; 
[sin(\theta+s) sin(\theta-s)]^{1/2}.
\label{CC.9}
\end{equation}
\\
The integral seems not to be analytically calculable, but it appears that a good approximation of it is $2sin\theta^{5/2}$.
Therefore, $\rho_j(\theta,0)$ reads
\\
\begin{equation}
\rho_j(\theta,0) \approx \frac{1}{\pi^3}sin\theta^{5/2}sin[(2j+1)\theta]\big].
\label{CC.10}
\end{equation}
\\
This expression is represented in Fig.~\ref{figureC3} for $j=0,1,2$ and 6 and is compared with the exact Wigner distribution.
The agreement is correct for $j=0$ and 1, good for $j=2$ and very good for $j=6$. 

\begin{figure}
 \includegraphics[width=200mm]{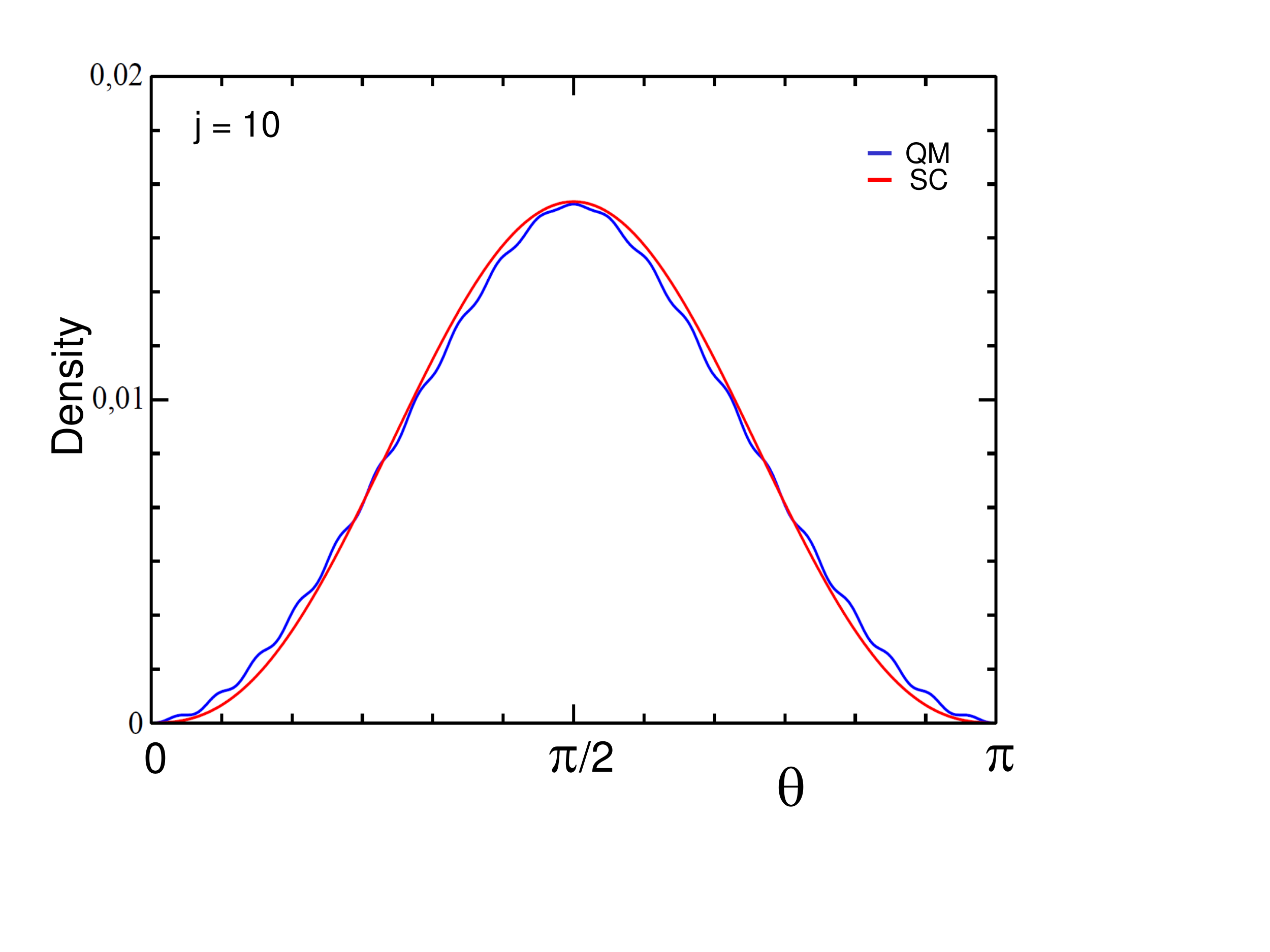}
 \caption{Comparison between the semiclassical value of $\rho_j(\theta,j+1/2)$ along the $D_3$ direction
 (see Fig.~\ref{figureC1}), obtained from Eq.~\eqref{CC.12}, and its exact quantum value for $j=10$.
 \label{figureC4}}
\end{figure}

Along direction D$_3$, defined by $P_{\theta}=j+1/2$ for sufficiently large $j$'s, we obtain from Eqs.~\eqref{CC.3} and~\eqref{CC.4}
\\
\begin{equation}
\begin{split}
\rho_j(\theta,j+1/2)
& = \frac{1}{4\pi^3}\;\int\;ds \; [sin(\theta+s) sin(\theta-s)]^{1/2} \\
& \quad + \left. \frac{1}{4\pi^3}\;\int\;ds \;e^{2i(2j+1)s}\; [sin(\theta+s) sin(\theta-s)]^{1/2}\right. \\
& \quad + \left. \frac{1}{2\pi^3}\;\int\;ds \;e^{i(2j+1)s}\; [sin(\theta+s) sin(\theta-s)]^{1/2}\; sin[(2j+1)\theta]\right..
\end{split}
\label{CC.11}
\end{equation}
\\
Since the second and third integrands strongly oscillates as compared to the first one, the above expression 
is well approximated by its first line. Using the previous approximation for the integral of the first line, we then arrive at
\\
\begin{equation}
\rho_j(\theta,j+1/2) \approx \frac{1}{2\pi^3}sin\theta^{5/2}.
\label{CC.12}
\end{equation}
\\
The quality of the agreement between this expression and the exact Wigner distribution is shown in Fig.~\ref{figureC4} for $j=10$.
 
Finally, the semiclassical expression of the rotational Wigner distribution along direction D$_4$, defined by $P_{\theta}=-(j+1/2)$,
is also given by the right-hand-side of Eq.~\eqref{CC.12}, due to the symmetry of the density with respect to D$_2$.

\newpage
%-------------------------------ACKNOWLEDGMENTS---------------------------------
\begin{acknowledgments}
%\section*{Acknowledgments}
L.B. is grateful to Profs. G. G. Balint-Kurti, A. Beswick and O. Roncero for valuable help regarding the derivations of
Appendix A. A.G.-V. acknowledges support 
by the Ministerio de Ciencia e Innovaci\'on, Spain, Grant No. FIS2010-18132, Consolider program, 
``Science and Applications of Ultrafast Ultraintense Lasers",   
Grant No. CSD2007-00013, and COST Action program CODECS, Grant No. 
CM1002. The Centro de Supercomputaci\'on de Galicia (CESGA), Spain,   
is acknowledged for the use of their resources.

\end{acknowledgments}

\newpage

%----------------------------------REFERENCES-----------------------------------

\newpage

 \clearpage

 \clearpage

 \clearpage

 \clearpage

 \clearpage

 \clearpage

 \clearpage

 \clearpage

 \clearpage

 \clearpage

 \clearpage

 \clearpage

 \clearpage


\begin{thebibliography}{10}
\bibitem{Schinke4}
Eqs.~(5.13) and (5.14) in ref.~\cite{Schinke} prove that for the harmonic oscillator, contour levels 
correspond to classical orbits.
\end{thebibliography}
\end{document}